\documentclass[a4paper,12pt,twoside,openright]{report}
\usepackage{ice-thesis-sjis}
\usepackage{url}
\usepackage{graphicx}
\usepackage{xcolor}
\usepackage{soul}
\usepackage{float}
\usepackage{CJKutf8}
\usepackage{amsmath}
\usepackage{bbm}
\usepackage[title]{appendix}
\usepackage[normalem]{ulem}
\usepackage{mathpazo}
\usepackage{multicol}
\usepackage{cite}
\usepackage{flafter}

\emergencystretch=\maxdimen
\makeindex

\title{Package Example}
\author{Team Learn ShareLaTeX}
\date{ }
	
\usepackage{longtable}
\newcommand\nomenclature[2]{#1 & #2 \\}
\usepackage{tocbibind} 
\usepackage{pdfpages}
	
\begin{document}
	
	\includepdf{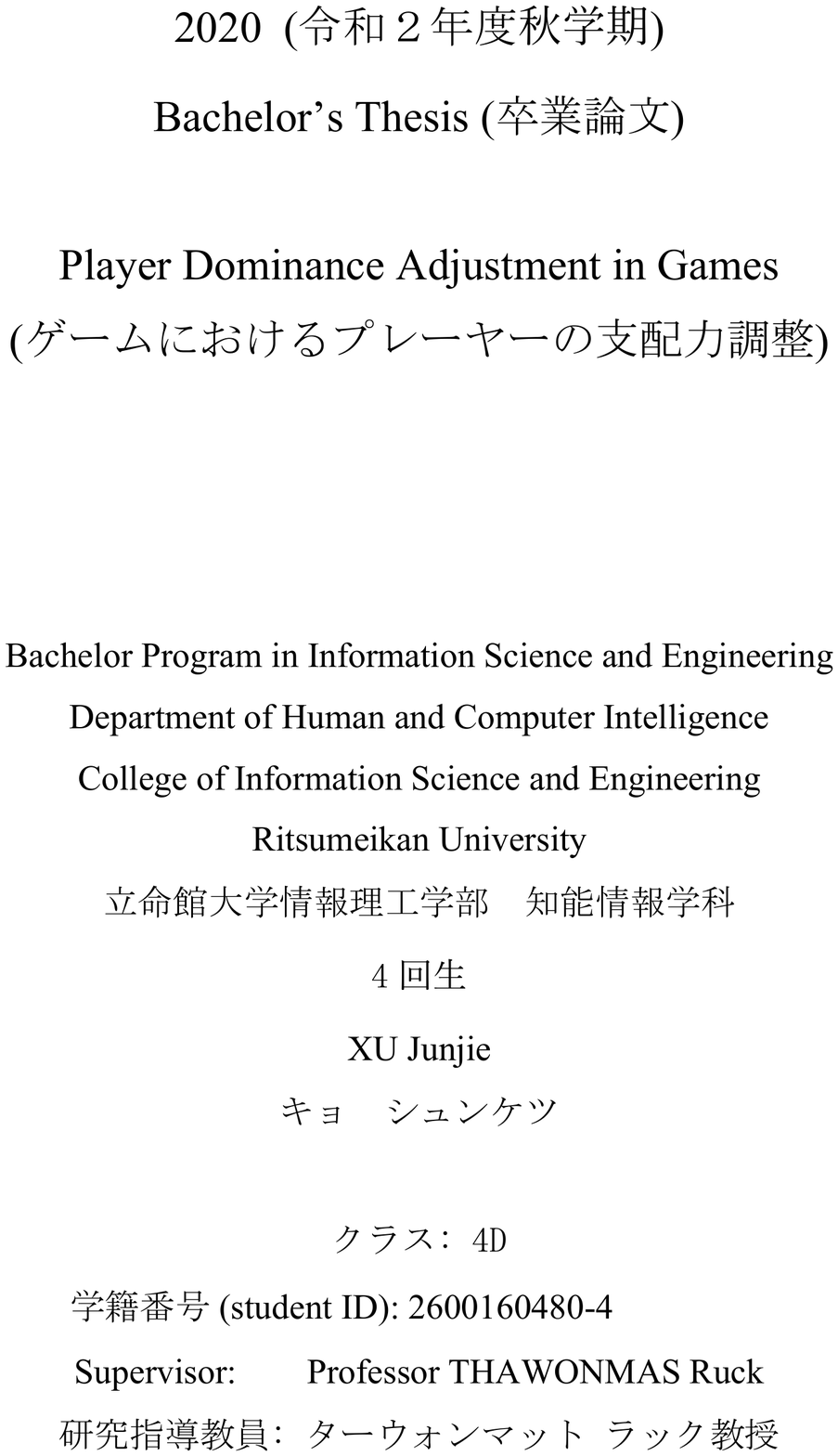}
	\newpage \thispagestyle{empty} \mbox{}	\newpage
	
	
	\pagenumbering{roman} 
	\parindent=11pt
	\noindent
	\begin{center}
		{\bf ABSTRACT}
		\vspace{1cm}\\
		{\bf Player Dominance Adjustment in Games}
	\end{center}
	\vspace{1cm}
	\par
	
Video Games are boring when they are too easy, and frustrating when they are too hard. In terms of providing game experience such as enjoyment to the player by match players with different levels of ability to player ability, We assume that implementing DDA for providing matches between player ability and overall game difficulty to the game, especially the modern game, has limitations in terms of increasing computational cost and complexities in the design of modeling the difficulty in modern games. 

To overcome limitations underlying the method of providing static difficulty changes to player, and DDA, we proposed a novel idea, ``Player Domination adjustment (PDA).'' The proposed idea is that to control the AI's actions based on the player's inputs so as to adjust the player's dominant power (e.g. the AI recognizes the player's attack actions but defends it in a wrong side to let the player incur damage to itself), which was proved as it leads to promotion of game-related self-efficacy in our work. Several pieces of research on were conducted on a social deduction game and a fighting game respectively, show our proposed idea has its potential of promoting User Experience(UX). As in an another study, outperforms DDA in two conducted experiments in terms of health promotion.

PDA also suitable in many cases of promoting the health of the player by increasing health metrics of the player such as balancedness ($Bal$, a term of in use of body parts in body movement). The mechanism is AIs or the whole game environment that selects next actions of AIs/ changes based on player's health metrics analyzed from his/her input action before such action is executed. With this mechanism we can stealthy encourage players performing the motions considering more healthy.

All experimental results demonstrate that PDA is capable of entertaining player, besides, benefits to the design of serious game which uses for health promotion. There is a promising direction for future research in PDA such as more investigations effects on other emotional factors, and for its better uses in various kinds of games.

	
	\newpage \thispagestyle{empty} \mbox{}	\newpage
	
	\begin{center}
		{\bf Acknowledgements}
	\end{center}
	\par
	Above all, I would like to thank my research supervisor, Prof. Ruck Thawonmas, who has supported me in all my endeavors and provided guidance, encouragement, and suggestion, gave me such many chances to publish papers in conferences organized in domestic or oversea. I appreciate his kindness to give his time generously. Besides, he has provided many useful materials as well as created an environment for me where it is a pleasure to study and conduct research.
	
	I would like to express special thanks to Dr. Pujana Paliyawan, who is like my second supervisor. I started my undergraduate research in advance with him and learned a lot of things necessary for writing papers, conducting experiments, polishing ideas for research, etc.. They helped me in all of my research studies, the paper would not be finished perfectly without their advice. I truly appreciate their time and kindness.
	
    Thanks also to my girlfriend Chen, for providing me consistent encouragement and caring within four years of undergraduate study, not only accompanies me in those wonderful lives, but also provides guidance for me in daily life even my research when required. Her existence keeping me optimistic when facing all the obstacles in my study and life in Japan. 
	
	Most importantly, I would also like to thank my parents for their love, support, and encouragement.
	
	Finally, Thanks to Ritsumeikan University and all the professors as well as lecturers who taught me in classes. I would never forget valuable experiences gained there for the instructions they gave.
		
	\vspace{50pt}
	\begin{flushright}
		\begin{tabular}{l}
			Xu Junjie\\
			\emph{Ritsumeikan University}\\
		\end{tabular}
	\end{flushright}

	\tableofcontents

	\listoffigures

	\listoftables
	\newpage \thispagestyle{empty} \mbox{}	\newpage
	
	
	\chapter*{List of Acronyms}
	\addcontentsline{toc}{chapter}{List of Acronyms} 
	\begin{longtable}{@{}p{3cm}@{}p{\textwidth \relax}@{}}
		\nomenclature{AI}{\underline{\textbf{A}}rtificial \underline{\textbf{I}}ntelligence, which is mainly referred to a non-player game character in this context}
		
		\nomenclature{$Bal$}{\textbf{\underline{Bal}}ancedness in use of body segment}%
		
		\nomenclature{$BF$}{\textbf{\underline{B}}alancedness \underline{\textbf{F}}itness}%
		
		\nomenclature{DDA}{\underline{\textbf{D}}ynamic \underline{\textbf{D}}ifficulty,  \underline{\textbf{A}}djustment}		
		
		\nomenclature{FightingICE}{The name of a \textbf{\underline{fighting}}-game platform developed by \textbf{\underline{ICE}} Lab\footnote{http://www.ice.ci.ritsumei.ac.jp/\~{}ftgaic/}}
	
		\nomenclature{GSE}{\underline{\textbf{G}}eneral \underline{\textbf{S}}elf- \underline{\textbf{E}}fficacy scale}	
		
		\nomenclature{GUESS}{\underline{\textbf{G}}ame \underline{\textbf{U}}ser \underline{\textbf{E}}xperience \underline{\textbf{S}}atisfaction \underline{\textbf{S}}cale}
		
		\nomenclature{ICE Lab}{\underline{\textbf{I}}ntelligent \underline{\textbf{C}}omputer \underline{\textbf{E}}ntertainment \textbf{\underline{Lab}}oratory, Ritsumeikan University }	
		
		\nomenclature{MCTS}{\textbf{\underline{M}}onte-\textbf{\underline{C}}arlo \textbf{\underline{T}}ree \textbf{\underline{S}}earch}
		
		\nomenclature{PDA}{\underline{\textbf{P}}layer \underline{\textbf{D}}ominance \underline{\textbf{A}}djustment}

		\nomenclature{PDR}{\underline{\textbf{P}}layer \underline{\textbf{D}}ominance \underline{\textbf{R}}ate}			
		
		\nomenclature{SE}{\underline{\textbf{S}}elf- \underline{\textbf{E}}fficacy scale}
		
		\nomenclature{UKI}{\textbf{\underline{U}}niversal \textbf{\underline{K}}inect-type-controller by \textbf{\underline{I}}CE Lab, an middleware which can convert detected motions into keyboard and/or mouse-click events, and send them as input to a targeted application.\footnote{https://sites.google.com/site/icelabuki}}	
		
		\nomenclature{UX}{\textbf{\underline{U}}ser \textbf{\underline{E}}xperience}		
		\nomenclature{Werewolf}{a social deduction game\footnote{http://aiwolf.org/}}

	\end{longtable}

\newpage \thispagestyle{empty} \mbox{}	\newpage

	\vspace{0pt plus 1fil}
	\pagebreak \setcounter{page}{1}
	\pagenumbering{arabic} 

\chapter{Introduction}

Video Game has a brief history, but developing at an ever-increasing speed, game console as well as PC video gaming has achieved a level of social and cultural influence. The widespread acceptance of video games over the past few decades has transformed the former niche hobby into a multi-billion dollar industry\cite{Kuo2017}. A recent report released by the Electronic Software Association \cite{Entertainment2019} illustrates the prevalence of gaming today: 75 percent of households in the USA have at least one video gamer in their household, and 65 percent of American adults play video games. Furthermore, gaming has begun to displace other forms of entertainment, with nearly half of all gamers preferring video games to movies and music \cite{Entertainment2019}.  Most of games are designed to entertain players, while another aspects of designing games is called serious game.

Regarding the first issue of entertaining players, because a UX of which the challenge level matches the skill of the human player is more entertaining than that is either too easy or too hard, a prominent research interest focus on providing even level of game satisfaction to players of different levels. There are two solutions to solving this issue. The first solution is allowing players to adjust static basic difficulty (easy, normal, hard) manually. As the second solution, introducing Dynamic Difficulty Adjustment (DDA)\cite{DDA}, a probabilistic model use for dynamically adjusting the difficulty of the game. However, regarding the first solution, the static difficulty leads to a lack of flexibility. Regarding the second solution, although this method also the ideally most effective to adjust game difficulty fits player ability, with the prosperity of the game industry, the increasing of game content in modern games leads that modeling as well as computing the game difficulty becomes a difficult task, especially for dynamic difficulty during gameplay.

Regarding the second issue of designing serious game for players, Serious games are video games that are created not for pure entertainment but for the purpose of solving issues in real world\cite{Serious2009} such as health or social issues.  In this thesis, we focus on the health problems, specifically, preventing the injuries by increasing a health metric called balancedness ($Bal$)\cite{pujanaAAAI, pujanaJournal}. Related to the literature, common causes of exercise damage are unbalanced exercise \cite{PPhillip} and repetitive movement \cite{EVOsite}. Unbalanced exercise is when one side of the body is more often used, leads to muscle imbalance, a cause of discomfort, injury, and physical ailments developing aches and pains \cite{PPhillip}. To prevent muscle imbalance, encourging balanced use ($Bal$) of segments on the two sides of the body, in other word, use a different muscle group frequently to allow muscles the opportunity to recover \cite{DSrinivasan}. We introduced a middleware application called UKI, to help players avoiding health risks by monitoring and enhancing a health metrics called balancedness ($Bal$) in use of body segments.

\section{Thesis approach}
To tackle the issues of providing entertaining game UX to player, or promoting health of player by enhancing health metrics. We proposed a noval idea  ``Player Domination adjustment (PDA)''. As a novel idea, we conducted several studys on Werewolf and FightingICE. 

In the case of Werewolf, we conduct the first pilot study using a famous social deduction game called Werewolf, using the concept of PDA\cite{myGCCE}, by letting human players play this game, one player at a time over the Internet. In the pilot-study setting, each subject thinks that he/she plays with six other players controlled manually by the backend of our system, following the specific rules based on PDA. Game situations are utilized to create two cases of gameplay: player-dominance games that most of the participant’s actions impact gameplay , non-player-dominance games that almost none of the participant’s actions impact gameplay. In this study, we evaluate the effectiveness by introducing General Self-Efficacy Scale (GSE)\cite{GSE} and Game User Experience Satisfaction Scale (GUESS)\cite{GUESS} as user evaluation. 

In this case of FightingICE, we present a fighting game AI as the player's opponent for promoting balancedness in the body parts' usage in full-body motion gaming by implementing PDA\cite{myMiG, myGPW}. For monitoring and enhancing $Bal$, we introduced a middleware application called UKI\cite{UKI}. Note that PDA is utilized to control the AI’s actions based on the player’s inputs in the way that adjusts the player’s dominant power. Namely, the AI analyzes an action that the player is going to perform and whether the action will increase the balancedness or not; if it does so, the AI  selects action which may give advantage to player, to let the player rule the game. On the opposite, it selects a strong action (obtained from Monte-Carlo Tree Search (MCTS)) in order to fight back to the player. As a comparison, this AI was compared with standard open-loop MCTS AI\cite{FTGMCTS} and an existing DDA AI designed for the same purpose\cite{kusanocog}.



\section{Thesis contributions}
This thesis contributes four major contributions as follow: 

\begin{enumerate}
	\item \textbf{Player Dominance Adjustment} A novel concept called ``Player Dominance Adjustment (PDA)''. , PDA is to manipulate the game process in the way that follows the player's intentions and make them feel that they have the power to dominate the game or that game situations go in the way they expect without adjusting the game difficulty. 
	
	\item \textbf{The Psychology Literture of PDA} As proved from the result in an pilot study for investigating the relationship between psychology effects and the variance of parameters in PDA such as Player Dominance Rate (PDR) conducted in our previous work\cite{myGCCE}, PDA can be used to promote game-related Self-Efficacy(SE) (the participant's belief that one can successfully execute the behavior required to produce certain outcomes) and as a result, enhance UX.

	\item \textbf{PDA for Entertainment} The result in an pilot study using Werewolf as testbed game, shows that increasing PDR in PDA in a simple way, could lead the enhancement in some aspects of player`s UX. Note that such enhancement of enjoyment in our previous study\cite{myGCCE} was not statistical large enough in this study.
	
	\item \textbf{PDA for Health Promotion} In our previous work, opposing fighting-game AI that designed for stealthy promotes the health of the player during motion gameplay was proposed\cite{myMiG, myGPW}. The AI decides its next action based on prediction on how each candidate action will induce the player to move, and how their health metrics will be affected. This AI received data representing the amount of movement on body parts of the player from UKI, and its goal is to promote the a health metric $Bal$, which represents the use of body parts. The results in these work show that the proposed AI outperforms the baseline as well as existing AI in promoting the $Bal$, providing a mechanism of enhancing health metric using PDA.
	\\
\end{enumerate}

\section{Thesis outline}
The remainder of this thesis is organized as follows: 

CHAPTER 2 provides literature reviews. This chapter contains three major sections, in the first section, we will describe the limitations in exciting methods for providing entertainment by matching player skills to the player. In the second section, we will focus on a psychology theory introduced by existing work between PDA and UX. In the last section of this chapter, we will mention what kinds of health factors can be increased by the use of PDA, and the middleware used for implementing our idea. Also, for health promotion, some of related work will be mentioned in this chapter.  

CHAPTER 3 introduces PDA. This chapter describes our motivation for establishing PDA, the definition of PDA, and provides an overview of the experiments conducted that successfully implemented PDA.

CHAPTER 4 introduces the details of our implementations. This chapter provides technical details on how to use PDA for the two purposes, enhance UX or for health promotion. We will also describe the details of the experimental setting of the conducted experiments as well as the results of those experiments.  

CHAPTER 5 describes the contributions, conclusions, and suggestions for future studies on the PDA of this thesis.

\chapter{Literature Review}

\section{Existing methods on game adaptation for players}

\subsection{Basic Static Difficulty Choices for Players}
\begin{figure}[htbp]
	\begin{center}
		\includegraphics[width=5cm]{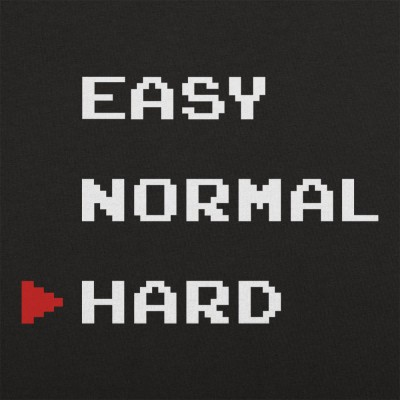}
		\caption{Static Difficulty Choices provided to players}
		\label{fig:difficulty}
	\end{center}
\end{figure}

Recently, The world of video games is rapidly and constantly evolving. The entertainment role of video games is predominant. However,  different players will have contrasting skills and motivations when playing games (e.g. the players without such high playing skill could not enjoy the game by experience too much failures due to the high difficulty, the expert players will feel boring when the challenges in the game is not enough). In many of the commercial games, as shown in figure \ref{fig:difficulty}, most popular approach by game developers is to introduce static levels of difficulty in their game (for example: easy, medium and hard) as its simplicity, to let the player choose and set their own difficulty. However, this approach does not take into account the player’s input in the game to vary the game’s challenge.

\subsection{Dynamic Difficulty Adjustment (DDA)}

DDA is the adjustment of changing elements in game, such as parameters, scenarios, and behaviors in a video game dynamically in real-time, based on the player’s ability. Commercially DDA has been implemented in a few cases\cite{GDC}, Hunicke introduced basic design for DDA\cite{DDA}. After that, There is a prominent research interest focus on providing even level of game satisfaction to players of different levels by DDA. DDA take into consideration the player’s input to balance the game’s difficulty dynamically. However, as the complexity of the modern games, DDA is difficult to implement, complicates tuning, to modify or measure difficulty in real-time is impossible task without an intensive design, sometimes even impractical.

\section{the psychology effects between Player Dominance Adjustment(PDA) and User Experience(UX)}

\subsection{Self-Efficacy(SE)}

As established in social cognitive theory\cite{cognitive}, Self-efficacy is defined as being either task specific or domain specific. Human motives and actions are governed extensively by forethought. The prime factor for influencing behavior is perceived self-efficacy, that is, people’s beliefs in their ability to perform a specific action required to achieve
a desired outcome. Self-efficacy is of a prospective and operative nature. perceived self-efficacy can be characterized as
being competence-based, prospective, and action-related\cite{cognitive}.

General self-efficacy is the belief in one’s competence to cope with a wide range of stressful or challenging demands, whereas specific self-efficacy is constrained to a particular task at hand, as established, also the proved effective questionnaire was  designed by Luszczynska et al.\cite{GSE} (see Fig. \ref{tab:SEori}).

\subsection{User Experience(UX)}

UX tells how the player experiences the gameplay and their satisfaction level. It was assessed based on six applicable factors of the Game User Experience Satisfaction Scale (GUESS) introduced by Phan et al. \cite{GUESS}, i.e., Usability, Engrossment, Enjoyment, Creative Freedom, Personal Gratification, and Visual Aesthetics. Each factor in use was measured by asking the player two questions; they are questions having the highest factor loadings in the original GUESS paper.

\subsection{Relationship between Game-related SE and UX}

\begin{figure}[htbp]
	\begin{center}
		\includegraphics[width=10cm]{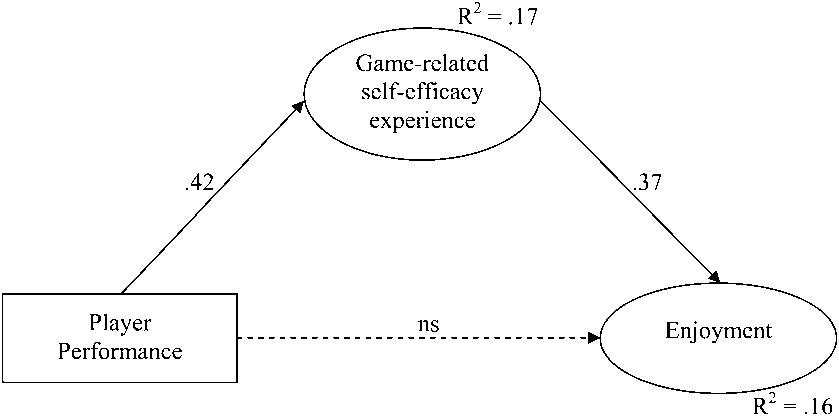}
		\caption{Structural equation model of the relationship among game related self-efficacy, player performance and enjoyment\cite{SEandUX}}
		\label{fig:SEUX}
	\end{center}
\end{figure}

Trepte et al.\cite{SEandUX} investigated the relationship between player performance, game-related self-efficacy experience, and game enjoyment (see Fig. \ref{fig:SEUX}). It's proved that self-efficacy is interrelated with enjoyment and that the relationship between game performance and enjoyment is driven by efficacy experiences.

Also, game related self-efficacy likely to be an important factor in understanding game enjoyment. Thus, not only the mere objective performance but also the user’s individual experience of self-accomplishment, competence, and control over the game environment constitute game enjoyment.

\section{Promoting healthy motion gameplay}
Despite the role of entertaining players, video games have been proven effective in providing exercise and rehabilitation \cite{GBarry,TBaranowski}. Games for health are not designed for replacing traditional exercises or sports, but for being a substitute for sedentary activities. Using many sources of exercises can help achieve physical activity guidelines.

\subsection{Balancedness ($Bal$) in use of body segments}
Balancedness in use of body parts is a key to muscle balance, a relative equality of muscle length and strength between opposite muscles. Muscle balance is vital for body movement and function \cite{PPhillip}. Muscles can become unbalanced by performing one-sided-type sports, or when requires a high level of physical activity with only one muscle or muscle group \cite{PMaffetone1}. Most cases of unbalanced muscle are unpreferable and considered dysfunctional \cite{PPhillip}. Muscle imbalance is a common cause of pain and should be prevented \cite{PPhillip,PMaffetone1}.

\subsection{UKI for motion gaming}

\begin{figure}[htbp]
	\begin{center}
		\includegraphics[width=1.0\textwidth]{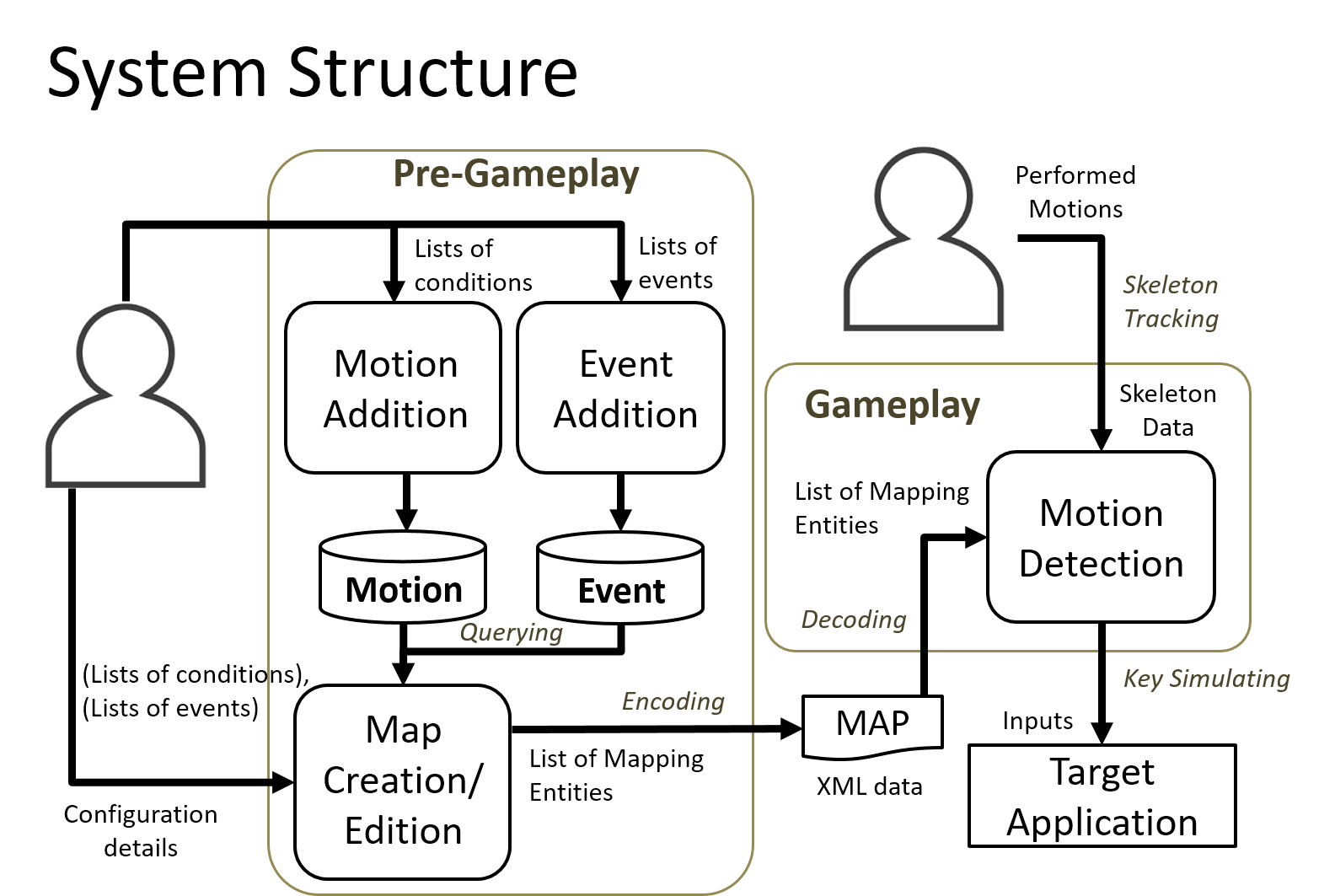}
		\caption{Overview of UKI\cite{UKI}}
		\label{fig:UKIOverview}
	\end{center}
\end{figure}

Here, as the purpose of monitoring health metrics of the player, we apply UKI\cite{UKI}(available in \cite{UKISite}), proposed by Paliyawan et. al, as a middleware for monitoring the amount of body movements and balancedness in use of body parts. The Universal Kinect-type-controller by ICE Lab allows its user to control any applications by using body motion as input. This middleware application was conceptually introduced in CIG and GCCE conferences in 2015 \cite{pGCCE2015,pCIG2015}, and the first completed version was presented in the journal of Software: Practice and Experience in 2017 \cite{UKI}. During gameplay, UKI receives streaming skeleton data from Kinect, translates detected motions into keyboard and/or mouse-clicks events and sends to a target application based on procedures predefined in a so-called MAP file (Fig.~\ref{fig:UKIOverview}). A MAP file is a file used to configure UKI for controlling a specific application. It consists of mapping components, describing how conditions are mapped to events.

{\bfseries Full-body Control:} Kinect has been recognized among gaming devices for its potential in providing full-body motion games for health promotion and rehabilitation. A systematic review on 109 articles has reported that research and development possibilities and future works with the Kinect for rehabilitation applications are extensive \cite{da2015motor}. The UKI project provides middleware that can facilitate the integration of full-body control with any existing games and applications \cite{UKI}. Besides such integration, UKI also has several features to enhance use, such as a module that allows the user to introduce new motions to the system by only performing them and a module for assessing calories consumption, balancedness in use of body segment, movement variability, and ergonomic health risks.

{\bfseries Health Assessment:} UKI's Health assessment module for FightingICE to calculate  the balancedness in use of body segments ($Bal$) by accumulating momentums of the player's body segments over time from the game starts. For calculating the momentum of body movement, this module uses Kinect captures 3D positions' raw data of 20 body joints. At first, coordination is localized to make data invariant to player's standing position. Joints on upper-body are centered to center coordination of shoulder, while those on lower-body are centered to center one of the hip. Secondly, the relative change between each pair of consecutive frames is computed by using Euclidean distance. Third, joints on the center of body are omitted, and remaining joints are grouped into four segments that are a pair of arms and a pair of legs. Details will be given in section 3.3. 

\begin{figure}[h]
  \centering
  \includegraphics[width=\linewidth]{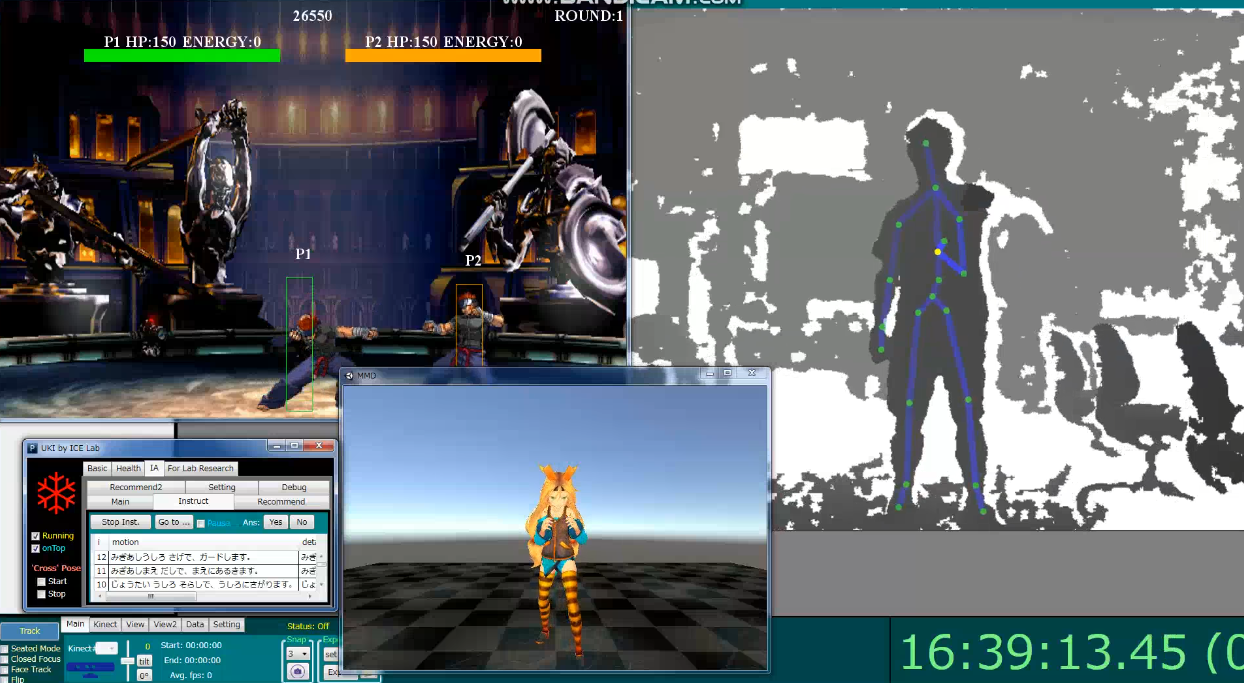}
  \caption{The visual instructor (the character in the middle bottom) is giving a tutorial to a player}
  \label{fig:instrctor}
\end{figure}

{\bfseries Pre-gameplay Instructor:}
Effective tutorial before gameplay is important for improving player engagement, especially for new players \cite{andersen}. This system applies to health-promoting motion games, as well as fighting game. As the example shows in Figure \ref{fig:instrctor},  we embed this module to guide the first-time player learning through the motions of controlling the game character in FightingICE, which enhances memorization and variety of the motions of the player would use 
\cite{IA}.

\begin{figure}[!tb]
	\begin{center}
		\includegraphics[width=1.0\textwidth]{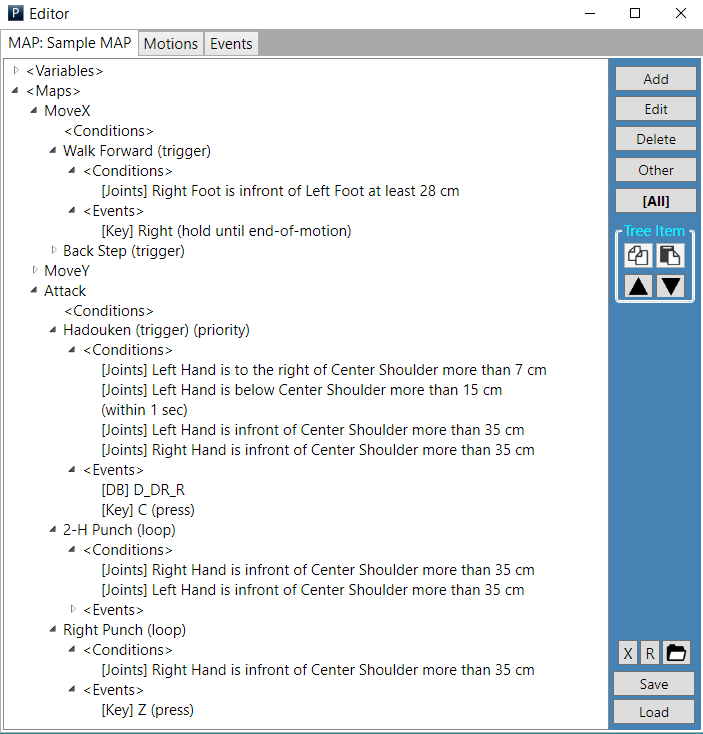}
		\caption{An example of a Map file\cite{UKI}}
		\label{fig:mapfile}
	\end{center}
\end{figure}

In addition, there is several advantages using this middleware UKI:

\begin{itemize}
    \item \textbf{Accessibility} : UKI uses a library of Microsoft Kinect SDK v1.8. It supports Windows 7 or above and thus covers most of Windows users nowaday. Also , it has user-friendly UIs, All UIs represents configuration details in natural language, allowing anyone to design conditions for detecting motions by only using Graphic UI without further programming or control with command lines.

	\item \textbf{Compatibility}: UKI works with any applications without accessing their source codes. The interactions is through sending simulated keyboard and/or mouse-click events from UKI to a target application, it works with applications regardless of their programming languages and platforms.

	\item \textbf{Flexibility}: UKI provides two types of motions: basic and sophisticated ones. Basic motions are single-step motions such as ``Right Punch,'' while sophisticated motions are those consisting of multiple steps, for example, ``Knifehand Strike'' can be performed by raise the right hand up, and chop the right hand down to the front of the body within a second after raising the right hand.
	
	\item \textbf{Usability}: Controls in games can be complex. There are cases where a sequence of key press inputs is intended to be sent without being interrupted, and cases where configuration changes is demanded on-the-fly during gameplay. The following are descriptions how UKI processes these cases. Some interruptions also can be handled by  an optional property ``Priority Process''\cite{UKI}. It uses an MAP file to control motions, which contains a list of mapping components for a certain application(cf. Fig.~\ref{fig:mapfile}). MAP file describes how to detect motions and what to do when they are detected,  and is loaded before gameplay. The problem is that there are cases where configurations are required to be changed on-the-fly during gameplay. For example, in Street Fighter, expected button combos for performing skills are subject to change corresponding to the player character’s facing direction. Such configurations capable of changing their contents during gameplay are called ``dynamic configurations''.

	\item \textbf{Facility}: the Motion Database and Event Database were added to allow the user to reuse a defined motion/event in several MAP files. Shortcut tools are also provided for detecting frequently used motion primitives, namely Atomic Postures (discussed in \hl{}3.3.1.3); the user can combine and/or change nuclear postures for detecting sophisticated motions. Mechanisms for monitoring of unhealthy postures, the amount of body movements, balancedness in use of body parts, and variability in movement are embedded to UKI. This middleware application was used in several health promotion studies\cite{refLiuAAAI, refGCCEishi, kusanocog}.
	
\end{itemize}

\subsection{HPAI--Health Promotion AI}

Paliyawan et al. \cite{pAAAI} introduced an adaptive motion gaming AI that induces its opponent, human player, to perform healthy motions. This AI uses historical gameplay data to generate a table of probability for predicting what counteraction the player likely to take when it performs a certain action. Nevertheless, the performance of this AI depends on prediction of the future counteraction of the player, which is hard to be accurate.

\begin{figure}[htbp]
	\begin{center}
		\includegraphics[width=0.8\textwidth]{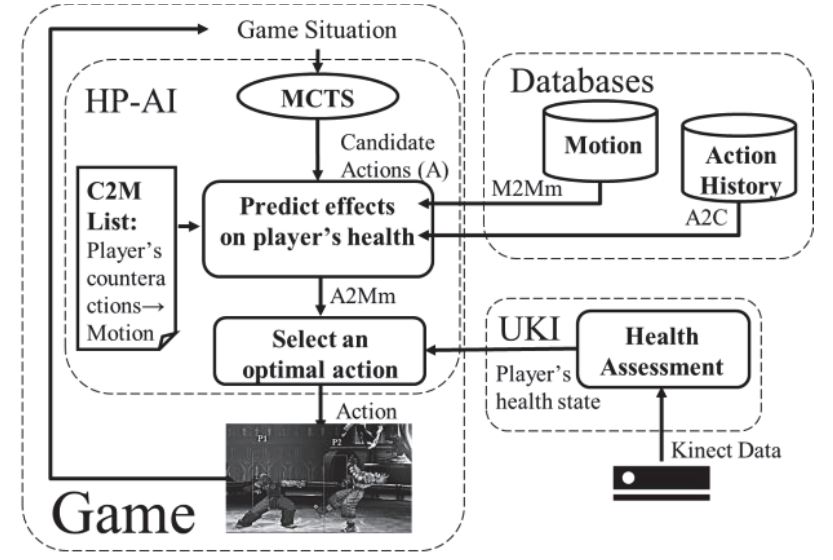}
		\caption{System Overview of HPAI\cite{pAAAI}}
		\label{fig:HPAI}
	\end{center}
\end{figure}

System overview for controlling the Health Promotion AI is shown in Fig. \ref{fig:HPAI}. The AI analyzes the player’s health metrics in real-time and uses supporting data from databases for determining its next move. Such determination gives the first consideration to improvement of the player’s health state, followed by strength of action. Derivation and information on computation of two essential data
used in the system: momentum of each segments of body movement and action-to-counteraction probability, was given in their work.

\subsection{DDAHP-AI--DDA Health Promotion AI}
\begin{figure}[htbp]
	\begin{center}
		\includegraphics[width=0.8\textwidth]{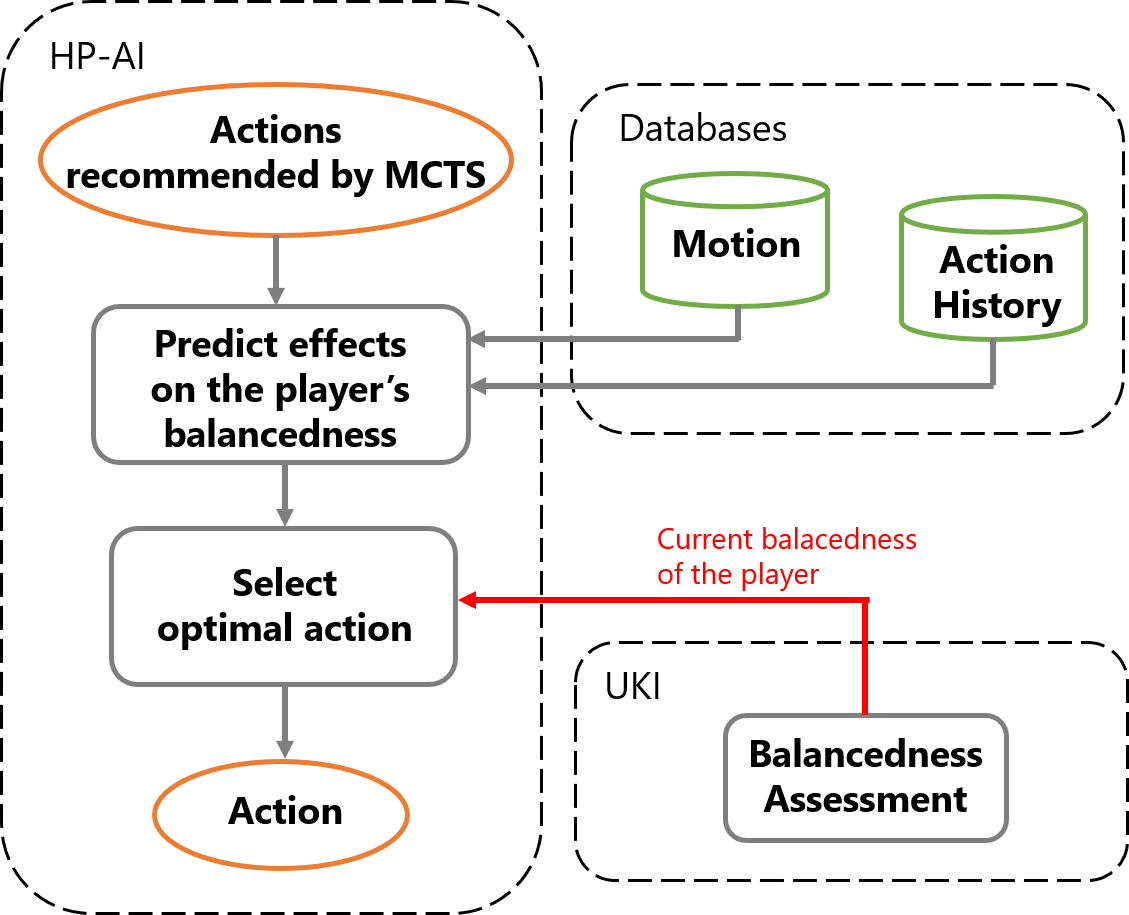}
		\caption{Overview of DDAHP-AI.\cite{kusanocog}}
		\label{fig:DDAHP-AI}
	\end{center}
\end{figure}

The improved version of this HPAI, DDAHP-AI \cite{kusanocog} was developed, the overview is given in Fig. \ref{fig:DDAHP-AI}, which with two addition mechanisms--use time series forecasting for more precise predictions and Dynamic Difficulty Adjustment (DDA) for better optimization between game skill level of player and difficulty of the game. 

As comparison to HPAI\cite{pAAAI}, this AI has some improvement in two aspects. First, it uses time series forecasting to more precisely forecast what actions the player will perform with respect to its candidate actions, based on which the amount of movement to be produced on each body part of the player against each of such candidates is derived; as the same as HPAI, this AI decides its action from those candidates with a goal of making the player's movement of their body parts on both sides equal. Second, this AI uses Monte-Carlo tree search that finds candidate actions according to DDA.

However, This AI has several limitations, such as (1) inaccurate prediction, especially with a small amount of history data, (2) fitting game difficulty to players with different game skill levels is difficult, and sometimes inaccurate \cite{AlexanderDDA}. DDAHP-AI has only little effects on promoting balancedness of player, which is also considered as the primary purpose of building these AIs.

\chapter{Player Dominance Adjustment (PDA)}

This chapter describes inspiration of establishing PDA, the definition of PDA, and provides an overview of the experiments conducted that successfully implemented PDA.

\section{Definition of PDA}

Player Dominance Adjustment (PDA), is defined to control the AI's actions based on the player's inputs in the way that adjusts the player's dominant power. Without adjusting the game difficulty, we manipulate the game process in the way that follows the player's intentions and make them feel that they have the power to dominate the game or that game situations go in the way they expect. 

\section{Pilot Study on a Social Deduction Game}

    We adopt PDA in Werewolf game\cite{myGCCE}, that can adjust its contents in the way that increases the player's dominant power and hence promotes self-efficacy and gameplay experience of the player. In this section, we describe a pilot study to test the PDA concept using a digital Werewolf game \cite{b6}, to evaluate the psychology effects, self-efficacy in this study, between PDA and UX.
    
\subsection{Motivation}
 
    The development of the Internet in daily life throughout the past several years and the immense success of video games foster the social competition of playing games\cite{b2}. If DDA is being used within the social competition of game playing, it is not difficult for players to notice when they rank themselves with other players. Comparisons under different game difficulties can lead to a negative perception by players, especially on their perception of achievement~\cite{b3}. Decisions made by the system to adjust difficulty contradicting the will of the players are possibly harmful to player experience~\cite{b4}.

    \subsection{Werewolf}
    Werewolf is a famous social deduction game. Each player in this game belongs to either the villager-side or the werewolf-side. The goal of human-side players is to eliminate all werewolf-side players, and the goal of werewolf-side players is to kill human-side players. The game alternates between night and day phases. At night, the werewolves secretly choose a villager to kill. During the day, the remaining villagers then vote on the player they suspect is a werewolf and kill that player (for details, see \footnote{https://www.youtube.com/watch?v=pehldN-JCPg}).

\subsection{Methodology}

	\begin{figure}[htbp]
		\begin{center}
			\includegraphics[width=12cm]{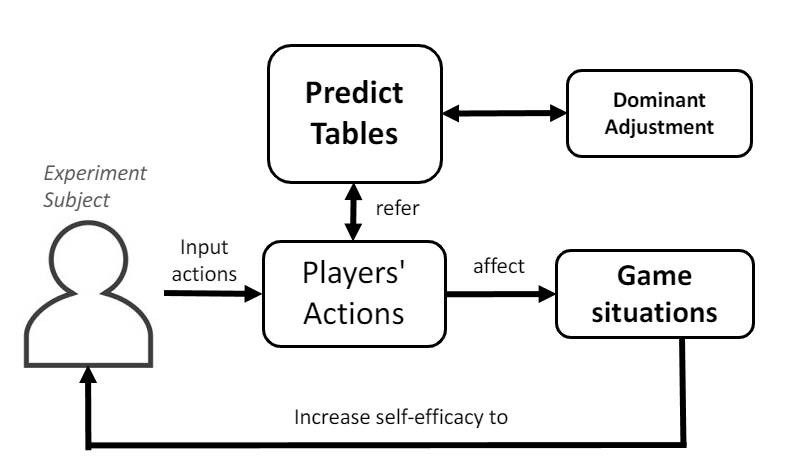}
			\caption{Overview of designed mechanism in digital Werewolf}
			\label{fig:WWoverview}
		\end{center}
	\end{figure}

\begin{figure}[htbp]
		\begin{center}
			\includegraphics[width=12cm]{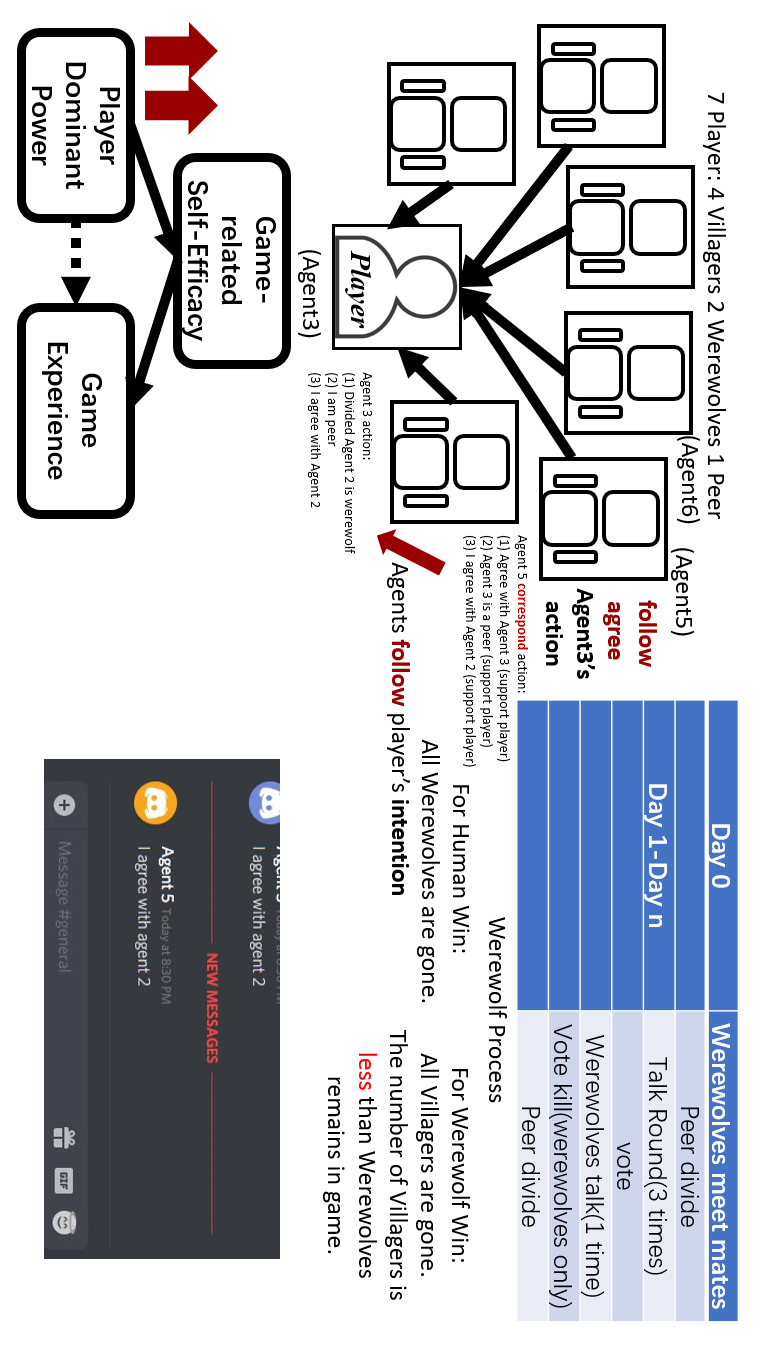}
			\caption{an example scene of player-dominance games in Werewolf, for top-left: player with 6 agents, for top-right: the rule of game in conducted experiment, for bottom-left: effect of PDA in theory, for bottom-right: live screenshot in player's screen in the conducted experiment}
			\label{fig:expWWpda}
		\end{center}
	\end{figure}

\begin{figure}[htbp]
		\begin{center}
			\includegraphics[width=12cm]{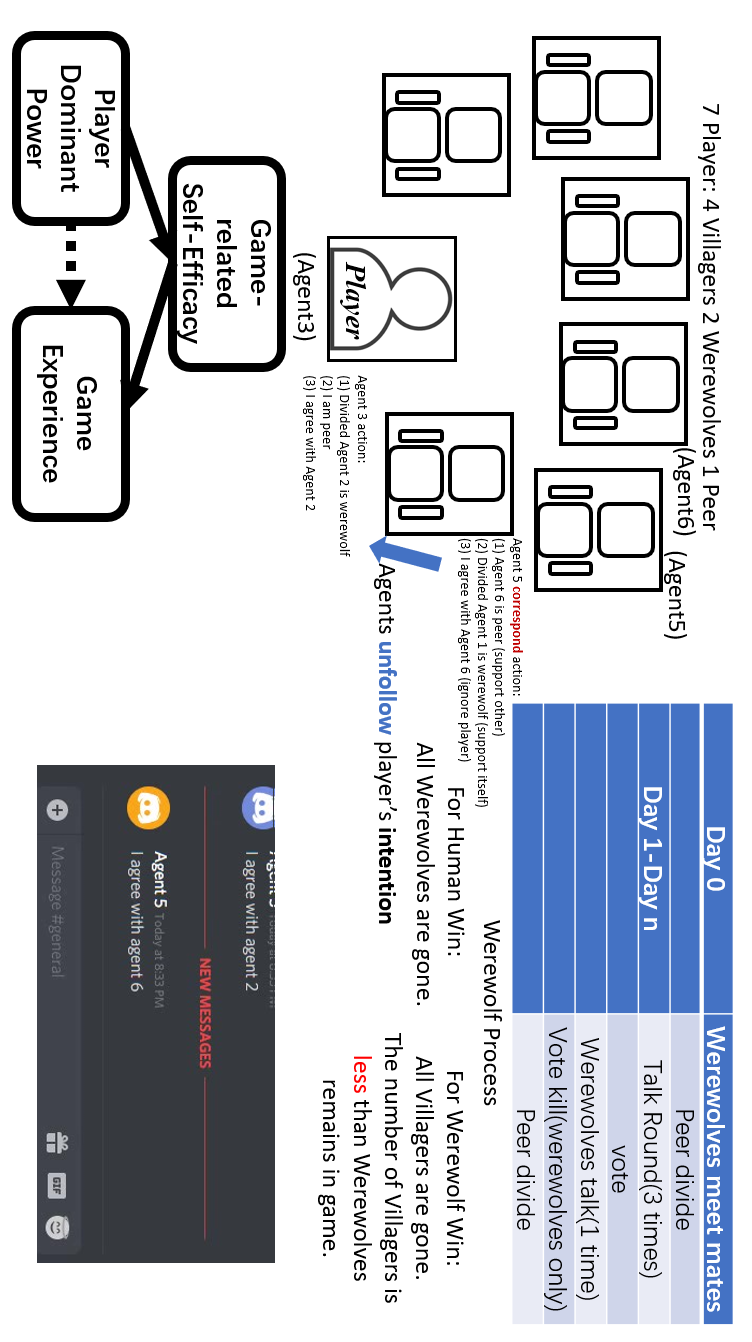}
			\caption{an example scene of non-player-dominance games in Werewolf, for top-left: player with 6 agents, for top-right: the rule of game in conducted experiment, for bottom-left: effect of PDA in theory, for bottom-right: live screenshot in player's screen in the conducted experiment}
			\label{fig:expWWnopda}
		\end{center}
	\end{figure}

\begin{figure}[htbp]
		\begin{center}
			\includegraphics[width=13cm]{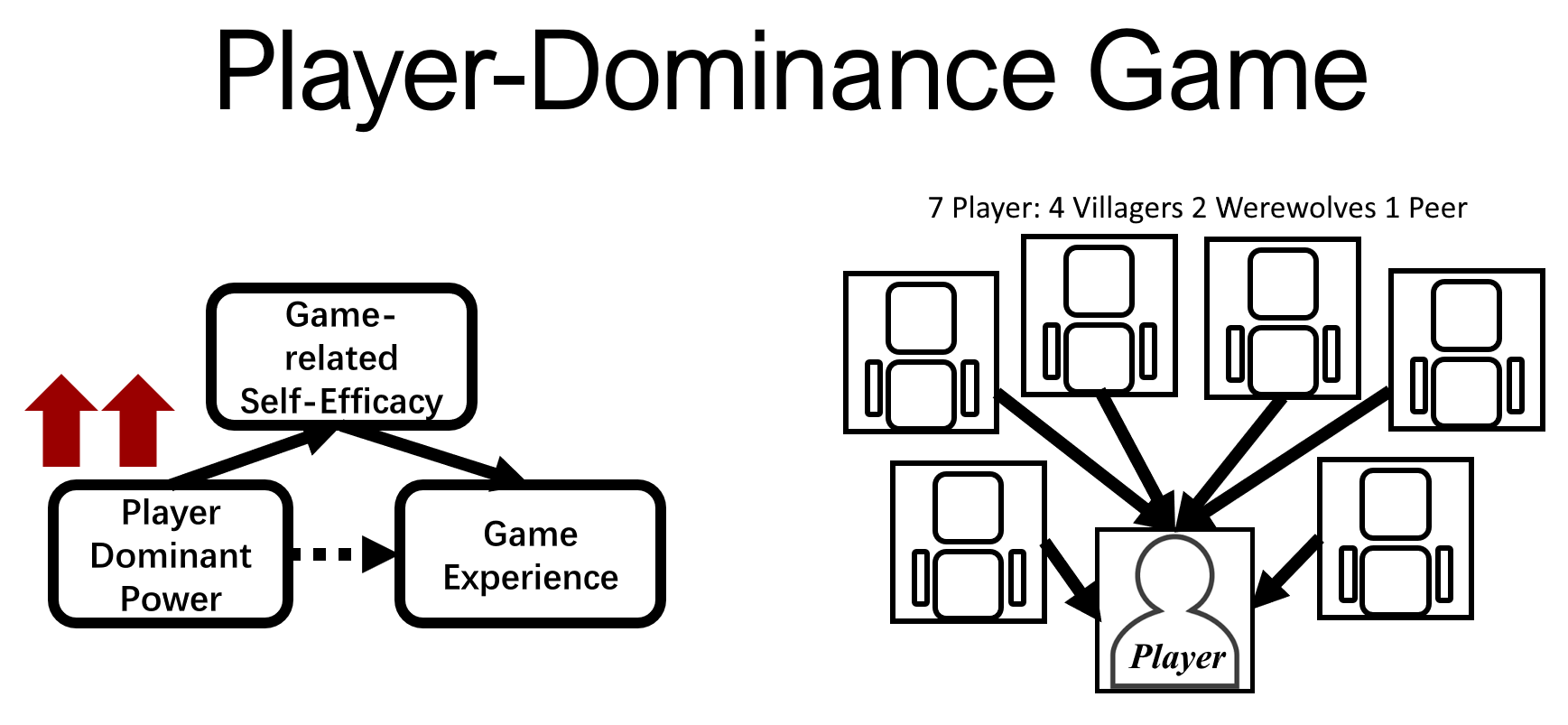}
			\caption{player-dominance games in Werewolf}
			\label{fig:WWpda}
		\end{center}
	\end{figure}

\begin{figure}[htbp]
		\begin{center}
			\includegraphics[width=13cm]{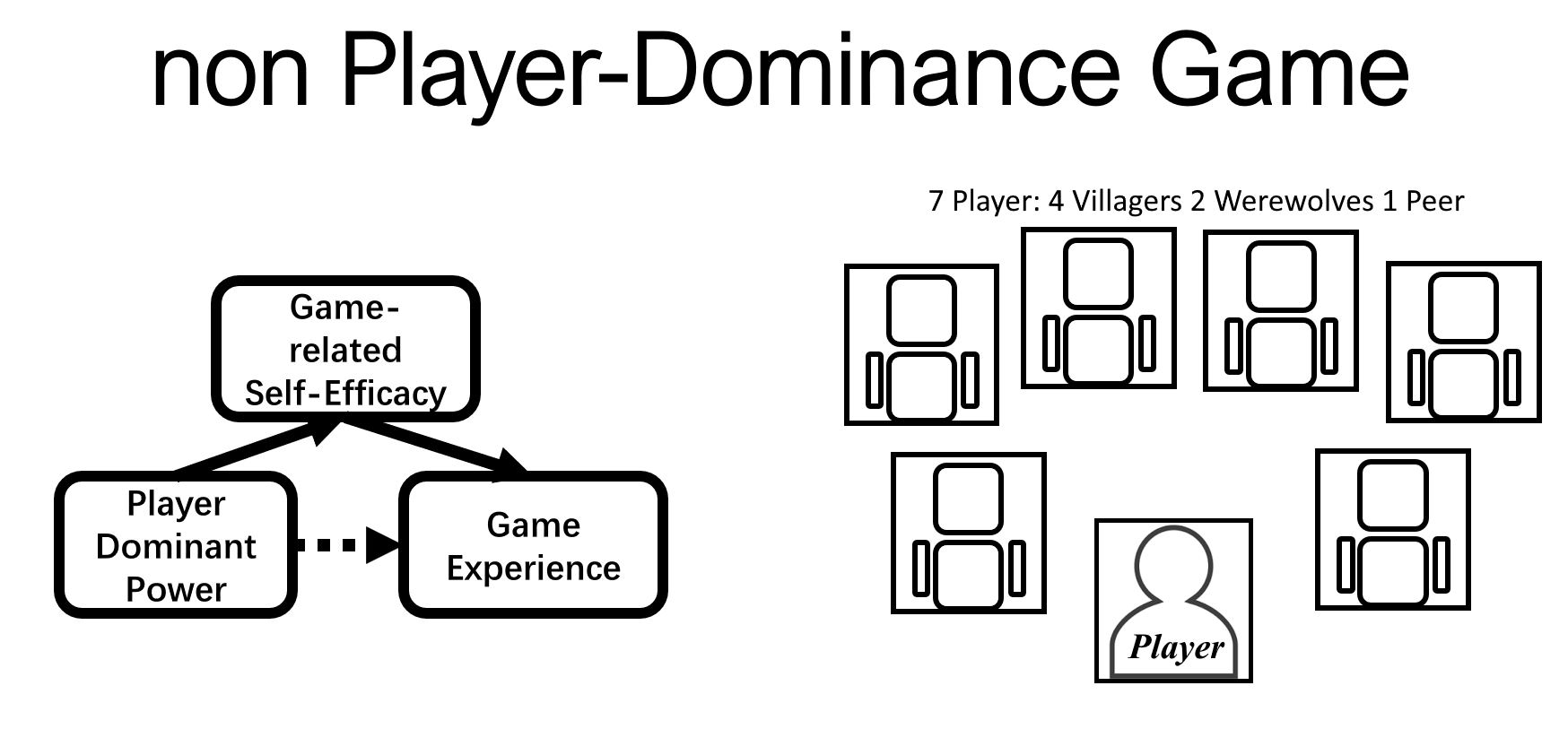}
			\caption{non-player-dominance games in Werewolf}
			\label{fig:WWnopda}
		\end{center}
	\end{figure}

    The overview of designed mechanism was given in Fig.\ref{fig:WWoverview}. By smartly adjusting this dominant power, we could increase game related self-efficacy of the player and enhance their experience of gameplay.
    The subjects were asked to play Werewolf game over the Internet under two cases of gameplay, player-dominance games(see fig. \ref{fig:WWpda}) and non-player-dominance games(see fig. \ref{fig:WWnopda}). In other word, one for letting the player has as much dominant power as possible as one for no dominant power at all vice versa. Specifically, for player-dominance games, the Agents would determine actions with prior consideration of player’s actions (as an example in fig. \ref{fig:expWWpda}). Inversely, for non-player-dominance games, the Agents would determine actions with less consideration of player’s actions(as an example in fig. \ref{fig:expWWnopda}). 
    
\section{Implementation on a Fighting Motion Game for Preventing Injures During Gameplay}

Motion games are widely used for health promotion as they can effectively motivate player engagement in physical activity during gameplay \cite{peng2013using}. Despite health benefits provided by motion games, we should also notice about effects, such as repetitive strain injury and muscle imbalance, caused by overusing some parts of body \cite{RRossler}. It is reported that muscle imbalance can be a primary cause of various aches and pains \cite{PMaffetone1}. 

As a solution to this problem, Paliyawan et al. \cite{pAAAI} developed an adaptive motion gaming AI that induces its opponent, human player, to perform healthy motions. This AI uses historical gameplay data to generate a table of probability for indicating what counteraction the player is likely to take when it performs a certain action. Nevertheless, the performance of this AI depends on prediction of the future counteraction of the player, which is hard to be accurate---further study by their group reported that the player's behavior change could suffer the effects of the AI \cite{kusanogcce}. In addition, although this AI induces the player to perform healthy motions, there is no reward encouraging the player to does so.

To overcome limitations underlying the above AI, we proposed a health promotion AI with PDA in this study\cite{myMiG, myGPW}. This AI tracks the amounts of movement on body parts of the player, and with these data, it determines whether a motion the player is trying to perform at a particular time is healthy or not in terms of promoting the balancedness in use of body parts. The AI encourages the player to perform healthy motions by making it easier for the player to hit the AI by actions related with those motions. Besides, the AI will take strong actions, obtained by Monte-Carlo Tree Search, towards the player's actions associated with unhealthy motions.

\subsection{Player Dominance Adjustment Health Promotion AI (PDAHP-AI)}
\begin{figure}[h]
  \centering
  \includegraphics[width=\linewidth]{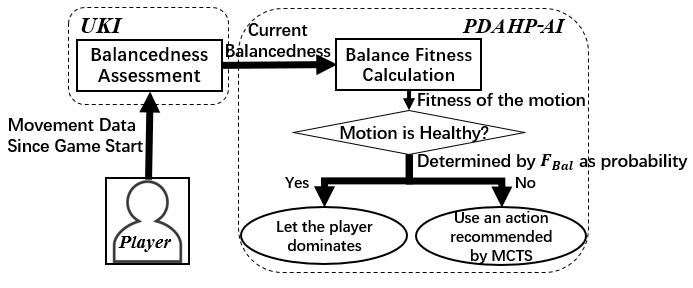}
  \caption{An overview of the system of PDAHP-AI}
  \label{fig:PDAAI}
\end{figure}

System overview for controlling the proposed health promotion AI (PDAHP-AI) is shown in Fig. \ref{fig:PDAAI}. If the motion of effective game skill is performed by player, the AI will analyze the player's input in real-time and uses motion movement table (M2Mm in Table \ref{tab:M2Mm}) \cite{pAAAI} to analyze the how much player's body segments move $am_{s}$ when he or she performs a certain motion which is going to be executed a game skill. Here, we induce a random number, which is ranged from 0 to 1 for decided action. If the amount of the number higher than $PDR$, the AI will determine action with strong AI, MctsAI in this paper. Adversely, the AI will determine harmless action but looks aggressive actions such as rush or walk toward the player that does no damage to the character's HP of the player.

During gameplay, the AI determines its next action based on prediction on how each candidate action will induce the player to move, and how their health will be affected. Candidate actions are recommended by a Monte-Carlo tree search (MCTS) algorithm, where in many AI competitions, MCTS-based AIs are ranked top. 
    
The calculated Player Dominance Rate $PDR$ remains until the next detection of player effective motion. The procedure will be executed when Kinect successfully detects the player's performed motion before it's going to be executed.

Namely, this AI is aimed at health promotion by introducing the concept of PDA, which are assumed to encourage the player to do healthy motion as we expected, by giving them chances to hit the AI and get positive feedback from gameplay or not, which also means adjusting their dominant power in gameplay. 

$BF$ calculated by the accumulated total movement and current movement of motion player performs, is used to calculated Player Dominance Rate $PDR$, the dynamic parameter to effect PDAHP-AI's determination of its action would be executed while player's skill is being executed. 

As the PDAHP-AI's overview in Figure \ref{fig:PDAAI}, for health promotion in FightingICE, the adjustment is based on analysis of the player's health parameters which is calculated by player's real-time input and, referred motion movement table and player's current health state to calculate the probability of whether situation should go as the participant expects or not, also called Player Dominance Rate in PDA, and its goal is to balance the movement amounts of body segments on the left side to those on the right side. If such motion is considered healthy, the AI will encourage the player to do such motion, by making it easier for the player to hit the AI by an action associated with that motion. On the other hand, the AI will counter actions associated with an unhealthy motion. In this figure, PDA does not need both measurement of player game skill level and fitting the game difficulty to the player and supposed to be a more effective way for reacting to player's action of health promotion rather than predicting player's action which has risk on imprecise prediction.

\clearpage

	As the comparison to previous AIs\cite{pAAAI, kusanocog}, there are three major benefits of this implementation as follows:
	
    \begin{itemize}
    \item {\textbf{No need to predict future counteraction}}: The AI directly receives information about the action the player is going to take from the middleware used for playing the game As it makes a decision based on this information, error from prediction is not an issue.
    \item {\textbf{Rewards to the player for performing healthy motions}}: Theoretically, when the player founds a certain action effective at a certain time (i.e., can hit the opponent), he will continue using it. We believe, making actions associated with healthy motion effective is to promote healthy motions.
    \item {\textbf{Better health with tailored game difficulty}}: We notice that MCTS-based AIs are usually very strong, especially in motion gaming. By the technique proposed, the difficulty will be adapted based on player's real-time inputs when the player performs healthy motion, which should make the player feels more fun. 
    \end{itemize}

\begin{table}
\caption{M2Mm (Motion to Movement momentum)}\label{tab:M2Mm}
\centering
\begin{tabular}{|l|c|c|c|c|}
\hline
Motion &  \multicolumn{4}{|c|}{Movement momentum}\\
\hline
 &  Right Arm & Left Arm & Right Leg & Left Leg \\
\hline
Right Punch & $5.83$ &$0.49$ & $0.51$ & $0.38$\\
\hline
Left Kick & $1.47$ &$1.68$ & $1.08$ & $6.42$\\
\hline
Crouch & $2.25$ &$2.11$ & $2.95$ & $3.04$\\
\hline
... &  & &&\\
\hline
\end{tabular}
\end{table}

\clearpage

\subsection{Monte-Carlo Tree Search (MCTS) for determining optimal action}

\begin{figure}[!ht]
	\begin{center}
		\includegraphics[width=13cm]{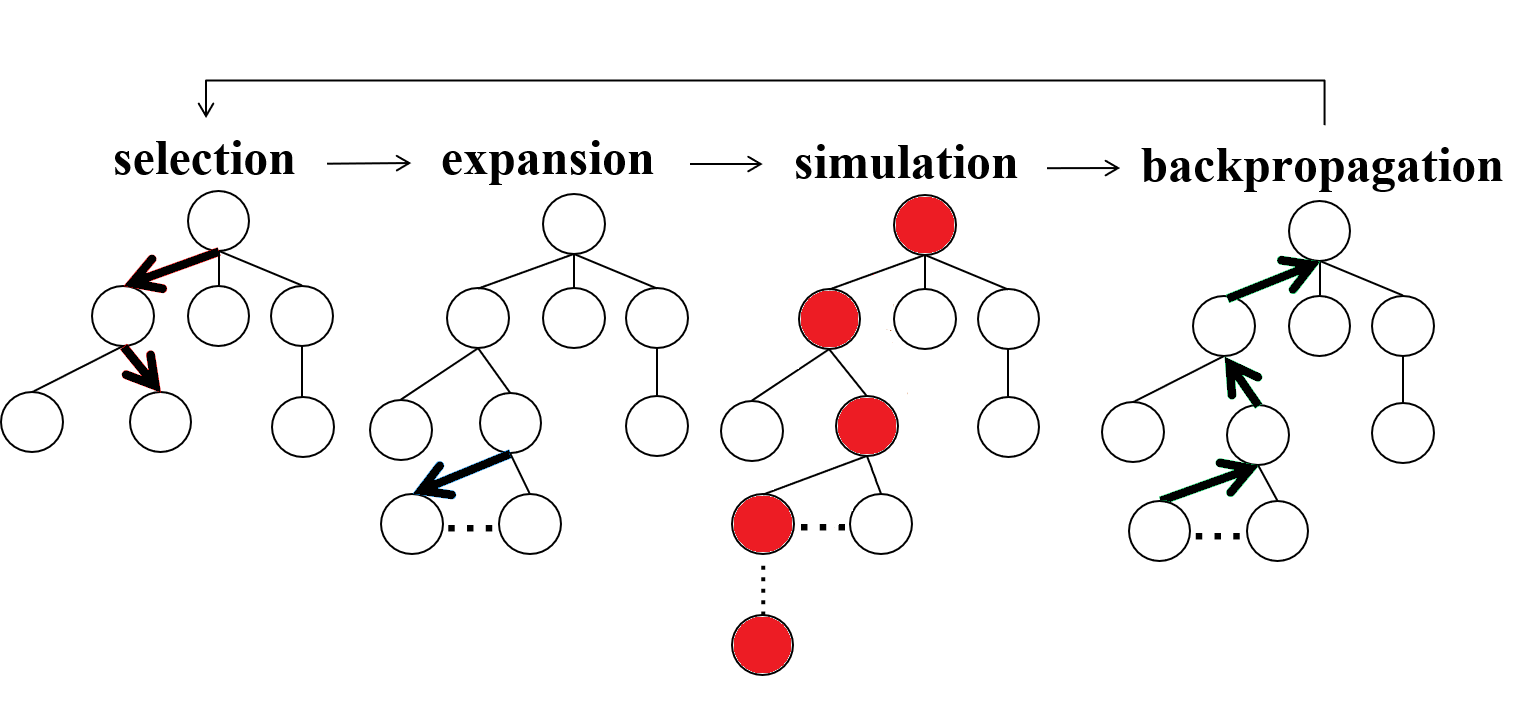}
		\caption{An overview of MCTS}
		\label{fig:mcts}
	\end{center}
\end{figure}

Monte-Carlo Tree Search (MCTS) module from MctsAi\cite{FTGMCTS} is embedded in our AI. In the MCTS process, four steps are repeated until preset fixed time $T_{max}$ runs out: selection, expansion, simulation, and back propagation. The description of each step is as follows (see Fig. \ref{fig:mcts}):

{\bfseries Selection}: Upper Confidence Bounds (UCB1) is introduced as the selection policy of nodes. The selection policy is given in Equation 3.1; considering the $i$-th node, $C$ is a balancing parameter, $N_i$ is the number of visits at that node. Reward used for evaluation is computed by using changes in hit points (HP) of player and AI before and after the actions is executed. $N$ is the times of visiting its parent node, and $\overline{X}_{i}$ is the average reward (see Equations 3.2 and 3.3).
\begin{equation}
  UCB1_i = \overline{X}_{i} + C\sqrt{\frac{2\ln{N}}{N_i}}
\end{equation}
\begin{equation}
  \overline{X}_{i} = \frac{1}{N_i}\sum _{j=1}^{N_i}{eval_j}
\end{equation}
\begin{eqnarray}
  eval_j = &\left(afterHP^{my}_{j}-beforeHP^{my}_{j}\right)& \nonumber \\ 
          &-\left(afterHP^{opp}_{j}-beforeHP^{opp}_{j}\right)&
\end{eqnarray}

{\bfseries Expansion}: After a leaf node is reached, if the depth of the path is lower than a fixed threshold and the number of visits of the leaf node is larger than a threshold, all adjacent child nodes will be created at once from the leaf node.

{\bfseries Simulation}: This step performs within a fixed time $T_{sim}$. At first, the AI will use a sequence of actions in the path from the root node to the current leaf node as AI actions, consequentially, it will perform actions randomly until reaching the same number as those in the path for the opponent's actions until reaches $T_{sim}$.

{\bfseries Backpropagation}: An update from simulation $eval_j$ is performed to obtain UCB1 for nodes that were traversed in the path. After reaching the end of the simulation, an update also performed for the UCB1 value of each tree node that was traversed in the path.

The fixed parameters used in the experiments are shown in Table \ref{tab:mctsparameter}. These parameters were set empirically through pre-experiments by previous work\cite{ishiharacog}.

\begin{table}[htbp] 
  \caption{Configuration of MctsAi's and MCTS module of PDAHP-AI} 
  \label{tab:mctsparameter}
  \hbox to\hsize{\hfil
  \begin{tabular}{ccl}
    \hline
    Notation & Meaning & Value \\
    \hline
    $C$ & Balancing Parameter & 0.42 \\
    $N_{max}$ &    Threshold of the number of visits & 7 \\
    $D_{max}$ &    Threshold of the tree depth & 3 \\
    $T_{sim}$ &    The number of simulations & 60 frames \\
    $T_{max}$ &    Execution time of MCTS & 16.5 ms \\
    \hline
  \end{tabular}\hfil}
\end{table}

\subsection{$Bal$: Balancedness}

\begin{figure}[htbp]
	\begin{center}
		\includegraphics[width=7cm]{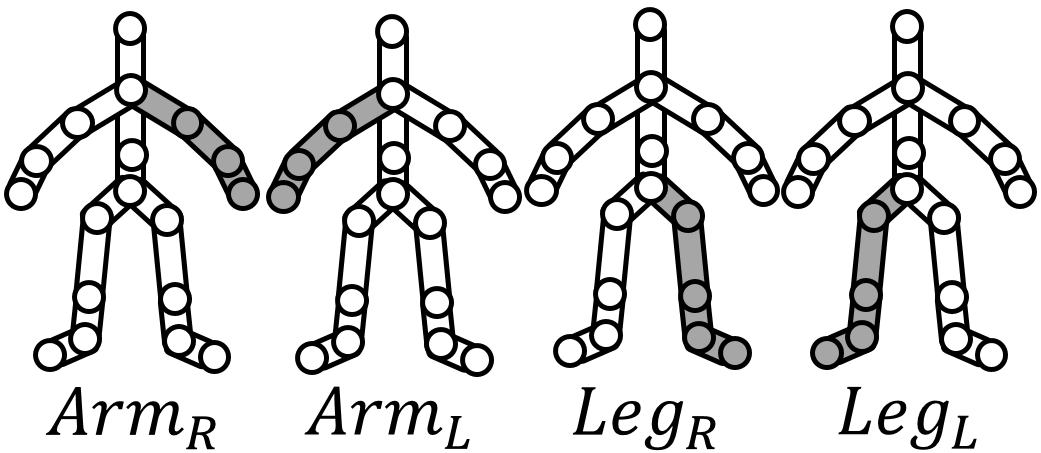}
		\caption{Four body segments for assessment of $Bal$.}
		\label{fig:aaaimm}
	\end{center}
\end{figure}

\begin{figure}[!ht]
	\begin{center}
		\includegraphics[width=13cm]{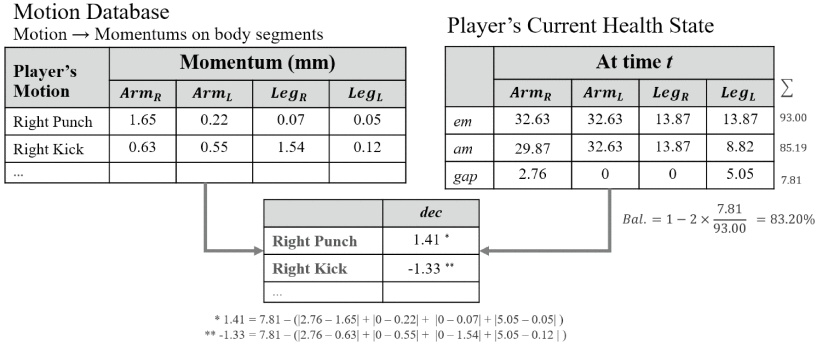}
		\caption{An example showing computation process of $Bal$.\cite{pAAAI}}
		\label{fig:aaaiprocess}
	\end{center}
\end{figure}

To measure $Bal$ \cite{pAAAI}(See example in Fig. \ref{fig:aaaiprocess}), a health metric uses in our study, we access values of the momentum of four body segments via UKI: Left Arm, Right Arm, Left Leg, and Right Leg (cf. Fig.~\ref{fig:aaaimm}). The momentum of a body segment corresponds an accumulated amount of movement of the segment since the beginning of a round (or when the player starts motion gameplay).  Let $am_s$ and $em_s$ denote the ground truth momentum and the expected momentum of the $s$th segment in the aforementioned four segments, respectively.

Changes of joints in the same segment are summed up to a change of segment. Finally, changes on a segment of interest over time are accumulated and represented by momentum in Equation 3.4. \cite{pAAAI}. 
\begin{equation}
  Bal = 1-2\times\frac{\sum _{s=1}^{4}{gap_{s}}}{\sum _{s=1}^{4}{em_{s}}}
\end{equation}

$gap_s$ (Equation 3.5) is the difference between the expected momentum $em_s$ and the actual momentum $am_s$ of the left or right segment in four segments of the body: Right Arm, Left Arm, Right Leg, Left Leg. Besides, $am_s$ is an accumulated total movement of the segment since each round of game starts, and $em_s$ is calculated in Equations 3.6 and 3.7.
\begin{equation}
  gap_{s} = em_{s} - am_{s}
\end{equation}
\begin{equation}
\begin{split}
  em_{RightArm} &= em_{LeftArm}\\
  &= max(am_{RightArm}, am_{LeftArm})
\end{split}
\end{equation}
\begin{equation}
\begin{split}
  em_{RightLeg} &= em_{LeftLeg}\\
  &= max(am_{RightLeg}, am_{LeftLeg})
\end{split}
\end{equation}

$Bal$ is used to compute the Player Dominance Rate to decide whether the opponent AI player should let the situation go as the participant expects or not.

\subsection{$BF$: Balancedness Fitness}

The term $BF$ describe in this section measures improvement in $Bal$\cite{pAAAI}. Specifically, it predicts a decrease in gaps between actual and expected momentum of the four body parts when a motion is performed. For each motion, UKI calculates this term by history data for predicting how performing it will influence $am_s$ as well as $em_s$ and $gap_s$ of the player.

The goal here of our study is use PDA to keep $Bal$ in a high value. Decrease in $gap_s$ leads to increase in Bal ; decx , computed by Equation 3.5, estimates decrease in $gap_s$ when the player performs the motion $x$, and in Equation 3.5, $mm_s(x)$ represents expected increase in the momentum of the $s$th body segment by this motion (i.e., an amount to be added to $am_s$ ). After $dec$ of all motions in use are obtained, their values are normalized into the scale of 0 to 1 by normalization, resulting in $F_{Bal(x)}$ in Equation 3.6.

\begin{align*} dec(x)=&\sum ^{4}_{s=1}{gap_{s}} - \sum ^{4}_{s=1}{|gap_{s} - mm_{s}(x)|} \tag{3.5}\\ F_{Bal}\left ({x}\right)=&\frac {dec(x) - dec_{min}}{dec_{max} - dec_{min}}\tag{3.6}\end{align*}

\chapter{Conducted Experiments}

This chapter describes the details of our applications and provides technical details on how to use PDA for the two purposes, enhance UX or for health promotion. The details of the experimental setting of the conducted experiments as well as the results of those experiments also would be discussed. 

\section{Pilot Study in Werewolf for enhancing UX}

\subsection{Apparatus}

This pilot study involved five university students as subjects, aged from 20 to 24 (4 males and 1 female). The subjects were asked to enter the pre-arranged online chat room and player Werewolf online on a PC. 

\subsection{Protocol}

First, the tutorial of playing our digital Werewolf in the chat room was given. the subjectsThen, they were asked about the experience of Werewolf gameplay before they attended this study. Each subject was asked to play Werewolf with six other agents over the Internet, which were actually controlled by the experimenter. Due to manipulation on voting and other game processes, gameplay can be generated in which most situations do not go as the subject expects (non-player-dominance game) and gameplay in which most situations go as the subject expects (player-dominance game). Each of the subjects experiences both cases, one each.

    \subsection{Evaluation Metrics}
    At the end of every gameplay, the subject did two self-reported questionnaires: General Self-Efficacy (GSE) scale \cite{GSE}(see Table \ref{tab:SEww}) and Game User Experience Satisfaction Scale (GUESS) \cite{GUESS}. GSE measures on Self-Efficacy on a 4-point Likert scale, while GUESS measures Enjoyment, Engrossment, Personal Gratification, and Playability on a 7-point Likert scale. The minimum value of the scales in both questionnaires is 1, the maximum scales is 7 vice versa.
    
    Regarding GUESS questionnaire, we derived 4 factors of the questions to our questionnaire. This questionnaire can be modified as game-related self-efficacy\cite{gameSE}, we modified it in our previous work, as in the use for Werewolf game(see Table \ref{tab:SEww})\cite{myGCCE}. These questions were translated into two other languages, Japanese and Chinese. We did evaluation on Enjoyment(Table \ref{tab:GUESS-EJ}), Playability(Table \ref{tab:GUESS-PA}), Play Engrossment(Table \ref{tab:GUESS-EG}) and Personal Gratification(Table \ref{tab:GUESS-PG}).

	\begin{table}[htbp]
	
		\begin{center}
			        \caption{\upshape{Items in factor Enjoyment, 7-point Likert Scale}}		
			\includegraphics[width=1.0\textwidth]{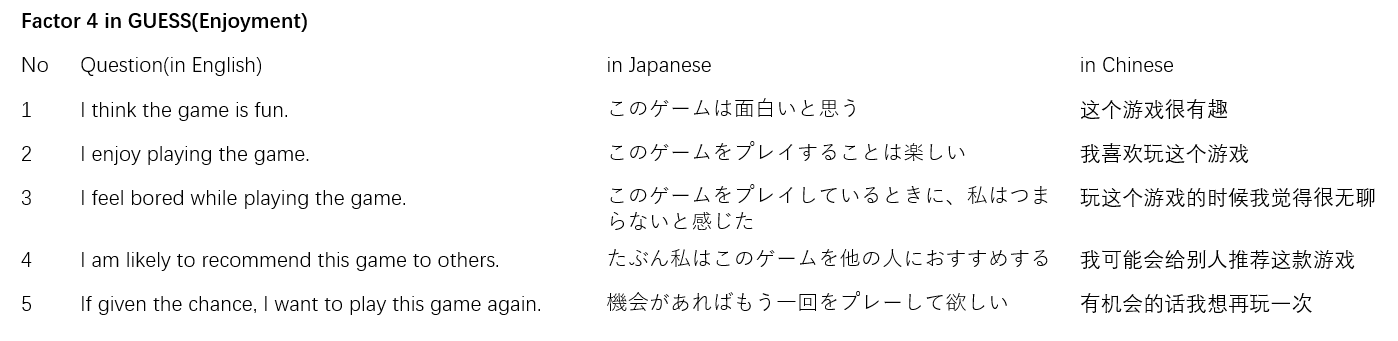}

			\label{tab:GUESS-EJ}
		\end{center}
	\end{table}

	\begin{table}[htbp]
		\begin{center}
			        \caption{\upshape{Items in factor Playability, 7-point Likert Scale}}		
			\includegraphics[width=1.0\textwidth]{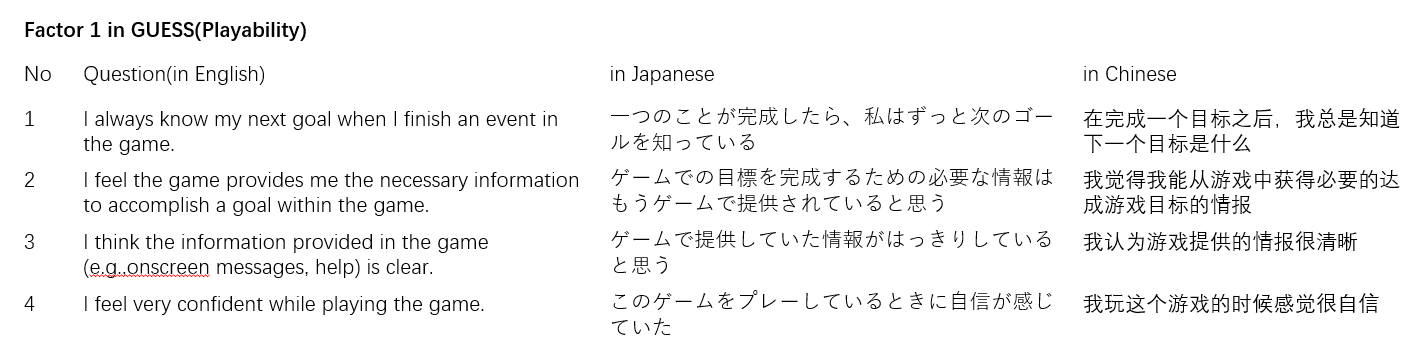}

			\label{tab:GUESS-PA}
		\end{center}
	\end{table}
	
		\begin{table}[htbp]
		\begin{center}
			        \caption{\upshape{Items in factor Play Engrossment, 7-point Likert Scale}}		
			\includegraphics[width=1.0\textwidth]{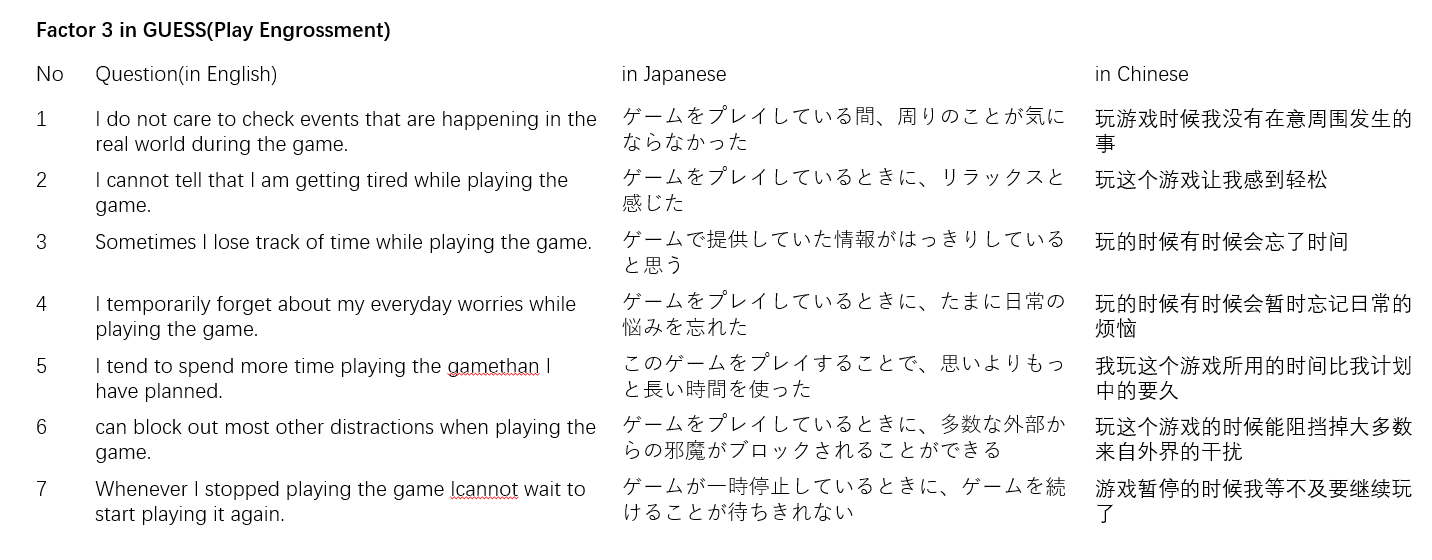}

			\label{tab:GUESS-EG}
		\end{center}
	\end{table}
	
		\begin{table}[htbp]
		\begin{center}
			        \caption{\upshape{Items in factor Personal Gratification, 7-point Likert Scale}}	
			\includegraphics[width=1.0\textwidth]{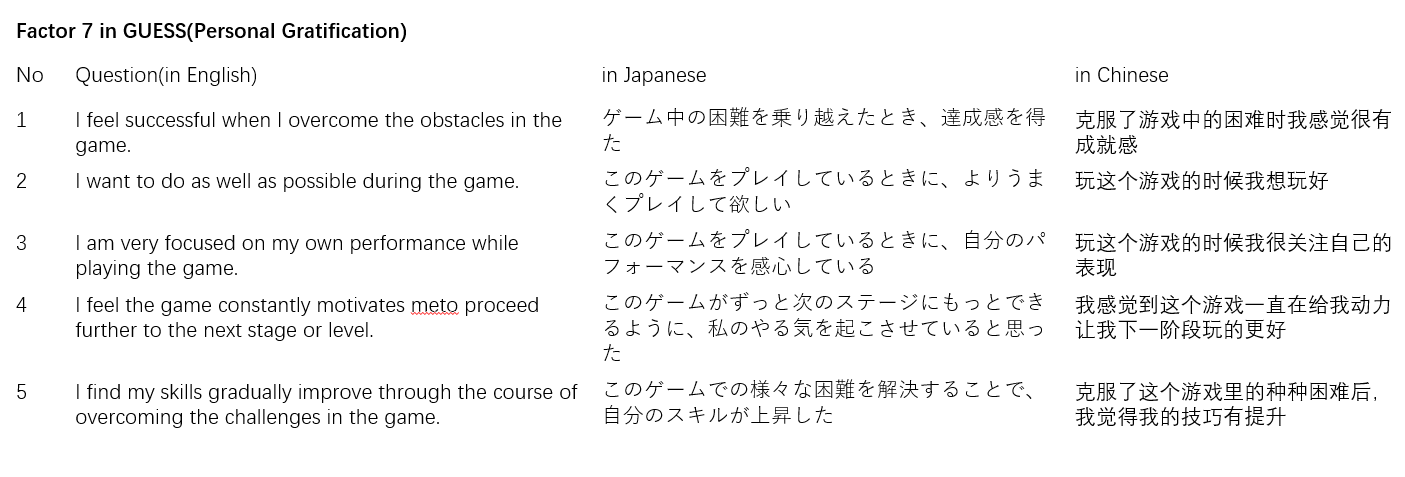}

			\label{tab:GUESS-PG}
		\end{center}
	\end{table}

	\begin{table}[htbp]
			\caption{\upshape{General self-efficacy questionnaire, 4-point Likert Scale\cite{SEtable}}}
			{\includegraphics[width=13cm]{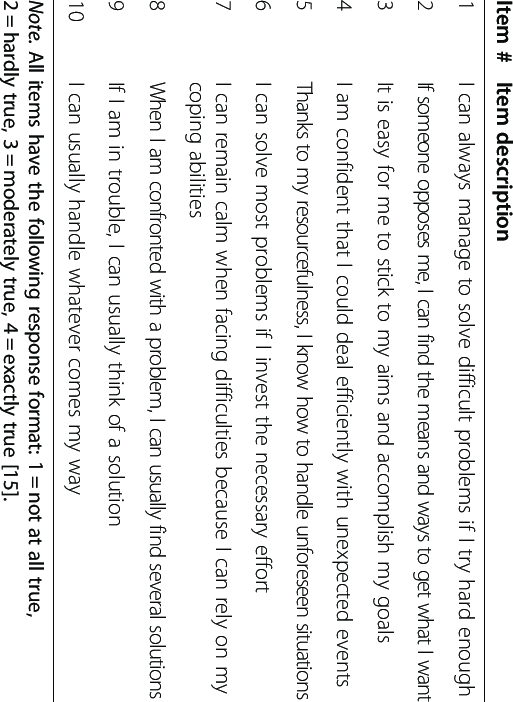}}

			\label{tab:SEori}

	\end{table}

	\begin{table}[htbp]
			\caption{\upshape{General self-efficacy questionnaire for Werewolf game, 4-point Likert Scale}}
			{\includegraphics[width=13cm]{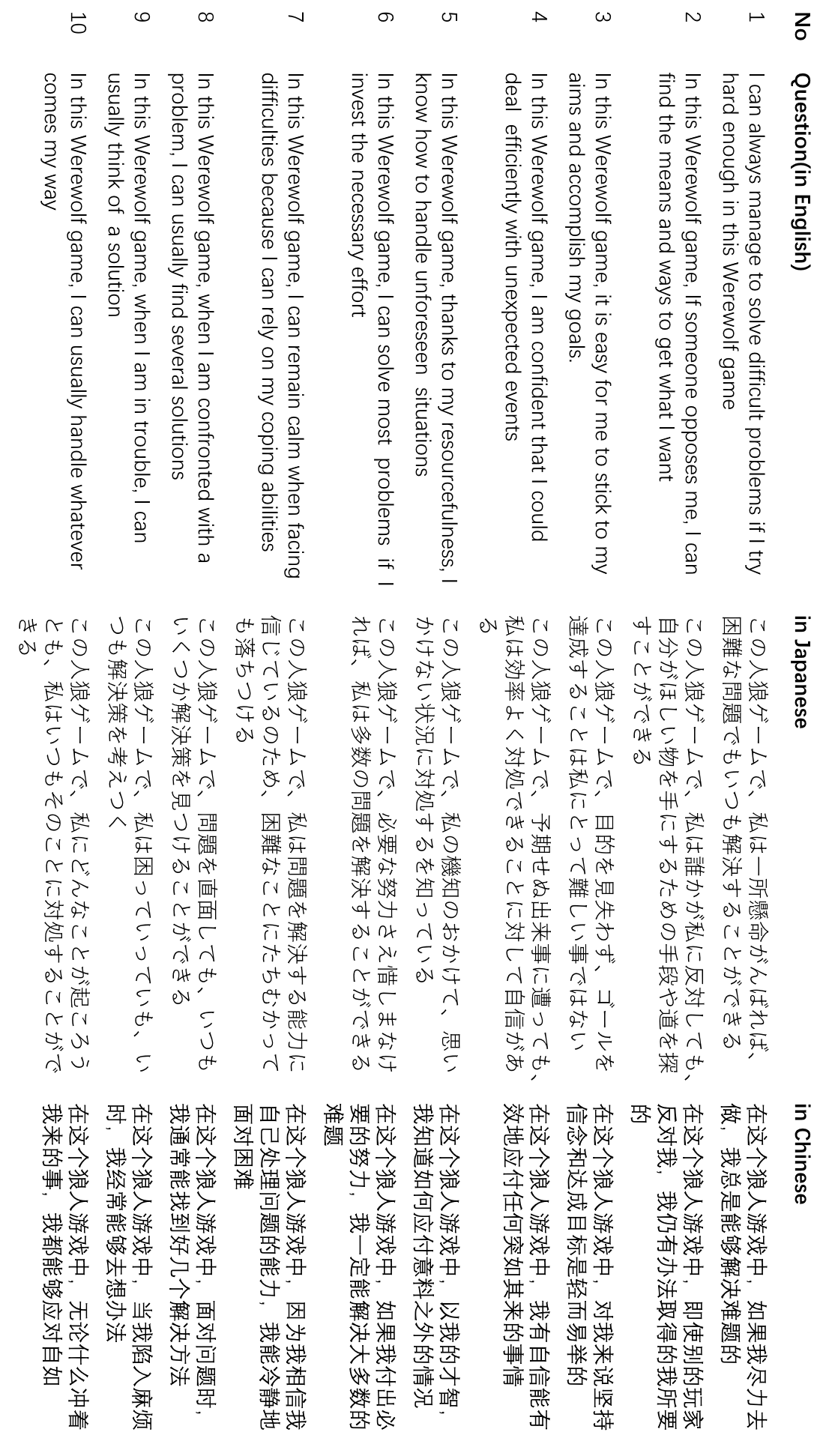}}			
			\label{tab:SEww}

	\end{table}

\subsection{Result}

Three subjects (subject 1, 2 and 3) are rookie players never playing this game or just several times, and two of them (subject 4 and 5) are experienced players to the Werewolf game, having experience of playing the game for more than two years.

\begin{figure}[htbp]
        \includegraphics[width=\linewidth]{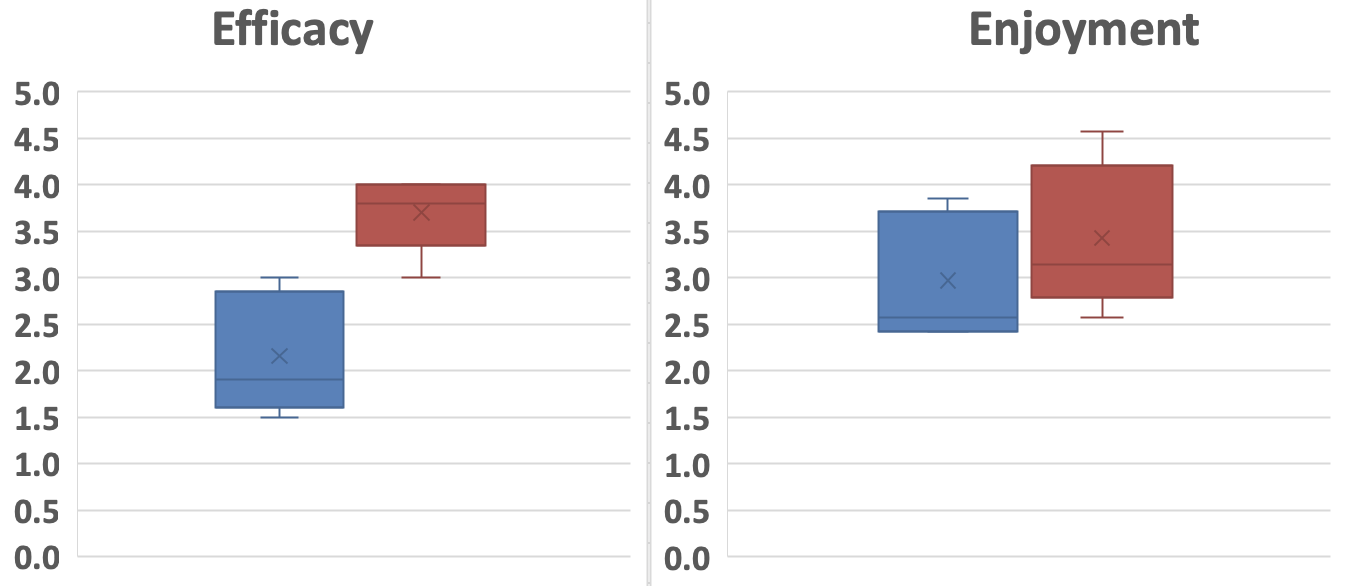}      
        \caption{he comparison Self-Efficacy and Enjoyment of the distribution between two cases of Werewolf gameplay (blue: case 0, red: case one)}

        \label{fig:WWdif}
    \end{figure}

    Comparing between two cases of gameplay using the Wilcoxon signed-rank test, there are statistically significant differences in Self-Efficacy ($p$-value: 0.041) and Playability ($p$-value: 0.043), but not in Enjoyment ($p$-value: 0.500), Engrossment ($p$-value: 0.336), nor Personal Gratification ($p$-value: 0.498). The comparison of the distribution between Self-Efficacy and Enjoyment was shown in Fig. \ref{fig:WWdif}.
    
    \begin{table}[htbp]
        \caption{\upshape{Self-Efficacy (SE), Playability (P), Enjoyment (Ej), Engrossment (Eg), and Personal Gratification (PG) of the five participants from the two cases.\cite{myGCCE}}}
        {\includegraphics[width=\linewidth]{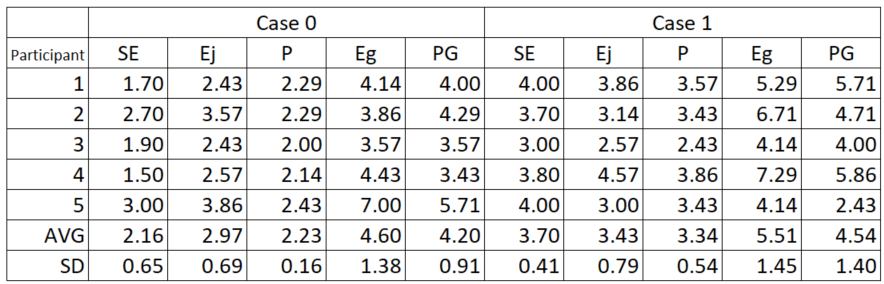}}
        \label{table:tab1}
    \end{table}

\begin{figure}[!ht]
	\begin{center}
		\includegraphics[width=13cm]{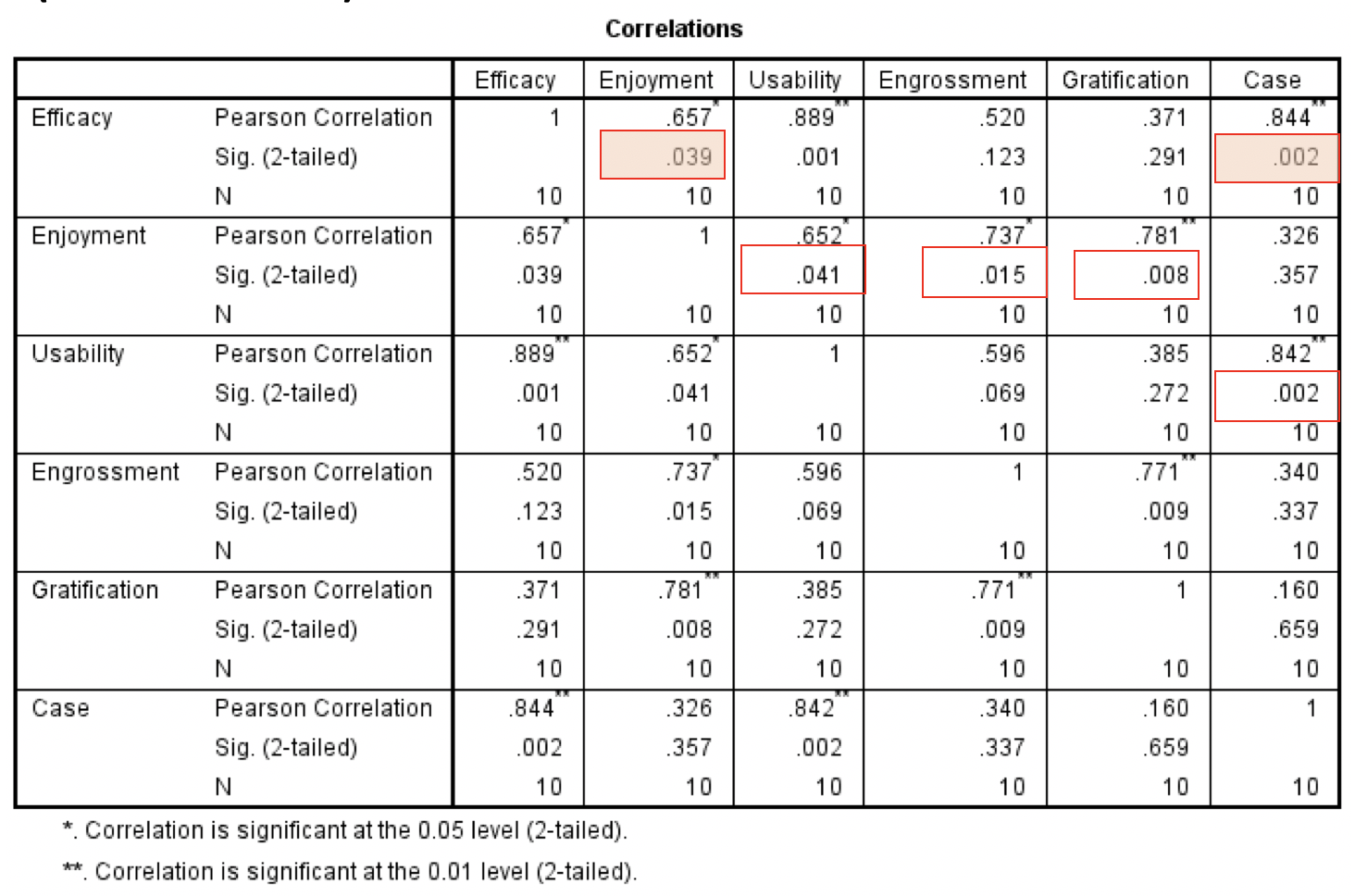}
		\caption{Pearson correlations between each pair of metrics of GUESS questionnaire in Werewolf experiment}
		\label{fig:WWpearson}
	\end{center}
\end{figure}

As in further investigation using Pearson correlations between each pair of metrics, a strong linear relationship is found between four pairs: (1) Self-Efficacy and Enjoyment, (2) Enjoyment and Playability, (3)Enjoyment and Engrossment, (4) Enjoyment and Personal Gratification (Fig. \ref{fig:WWpearson}).

    \clearpage

\subsection{Conclusions}
    This study shows that stealthy increasing dominant power of the player significantly increased Self-Efficacy. In addition, although Enjoyment was not higher in player-dominance games in this pilot study, Enjoyment still had a strong linear relationship with Self-Efficacy. Therefore, we find that enhancing dominant power of the player indirectly increase Enjoyment. Out of our expectation, such increases might not be large enough in this study, also, the difference between the two cases was not statistically significant. We assume the significant change between the two games is a main reason leading to the result that the enjoyment did not significantly increase as we expected.

    \clearpage

\section{PDAHP-AI for Health Promotion}

\begin{figure}[!ht]
	\begin{center}
		\includegraphics[width=1\linewidth]{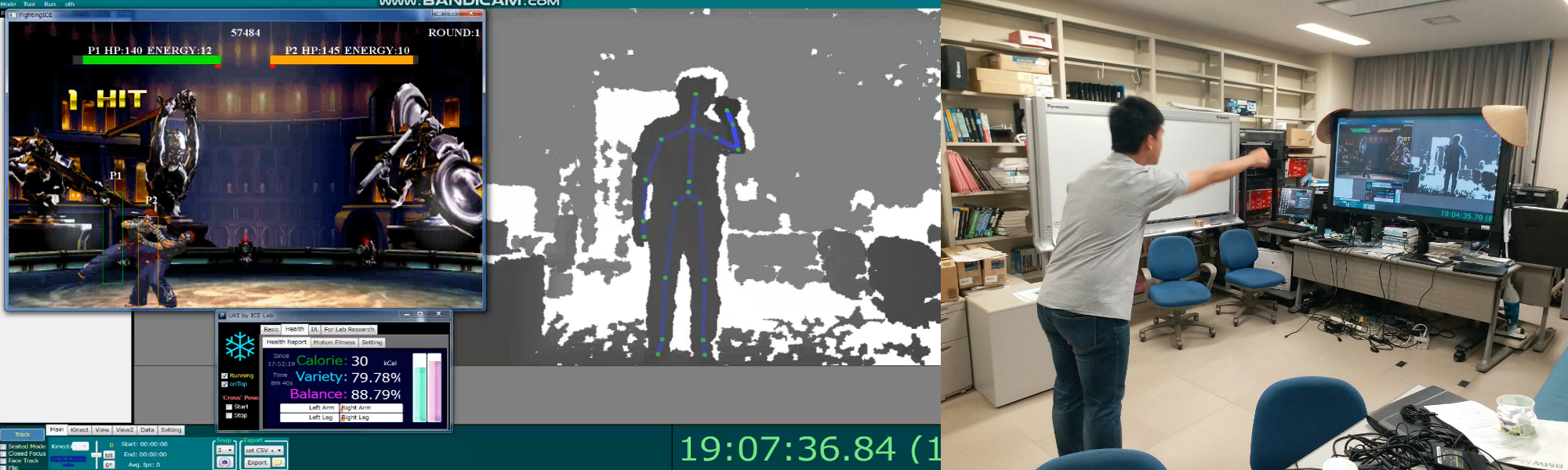}
		\caption{(Left) a screenshot of FightingICE and (Right) a player playing FightingICE}
		\label{fig:screenshot}
	\end{center}
\end{figure}

The proposed AI\cite{myGPW, myMiG} is implemented on FightingICE \footnote{http://www.ice.ci.ritsumei.ac.jp/~ftgaic/}, a fighting game platform for AI development and competition which was used to hold annual AI competitions at the IEEE conference on Computational Intelligence and Games (CIG) during 2014-2018 and  Conference on Games (CoG) since 2019. UKI \cite{UKI}\footnote{https://sites.google.com/site/icelabuki/} is used for integrating full-body control with the game, as well as assessing the amount of body movement and health parameters of the player (demo video is available\footnote{http://tiny.cc/as839y}). Experiment on 18 university students is conducted to evaluate the proposed AI in health promotion and game difficult. The AI is compared with a typical open-loop MCTS AI, named MctsAI, by Yoshida et al. \cite{FTGMCTS}, in terms of health promotion and game difficult. 

\subsection{Apparatus}
The application used are FightingICE (Java) and UKI (C\#) and ran simultaneously, and a screen during gameplay is shown in Fig.~\ref{fig:screenshot}. The player stood within a range between 2 and 4 meters from a 65-inch LCD screen (Panasonic TH-65PB2J) with the resolution of 1920 x 1080.

\subsection{Protocol}
We recruited 18 university students to participate in the experiment in the lab, from first-year bachelor's degree to graduate students, who are totally new to FightingICE. Before playing two rounds of the game for the main experiment, each participant was trained by a pre-gameplay virtual instructor. This instructor guides the participant to all motions available for game control and ensure that the participant can perform the motions correctly. After instruction, players in group 1 played the game against MctsAI, and then against PDAHP-AI. For players in group 2, they played the game against PDAHP-AI, and then MctsAI.

\subsection{Result}

\subsubsection{Comparison with MctsAI}

Comparing between fighting against MctsAI and PDA-HPAI, a significant difference in $Bal$ at the end of gameplay was confirmed by paired sample t-test ($p$-value = .033) and Wilcoxon signed-rank test ($p$-value = .031). The average $Bal$ is higher when the player fighting against PDA-HPAI (Table \ref{table1}), indicating that the proposed AI works as our expectation.

$HpDiff$, the difference of hit point (HP) between player and AI, is calculated at the end of each round by subtracting HP of the player character by HP of the opponent AI character. As the initial HP is 150, and each round ends when the player or the AI is defeated, therefore values of $HpDiff$ range between -150 and 150, where the negative values indicate that the player loses while the positive values indicate that the player wins. There is no significant difference in $HpDiff$ when fighting against different AIs based on statistical tests. However, the average $HpDiff$ is higher when the player fighting against PDA-HPAI (Table \ref{table1}). Besides, all the participants win a total of 2 rounds against MctsAi and 7 rounds against PDA-HPAI that also means the challenge playing against PDA-HPAI is more close to the ability of players, considered to be more fun based on Flow Theory \cite{flow}. Moreover, the remaining HP is closer when the player faces PDA-HPAI then faces MctsAI, that means the proposed AI is more likely to ensures that players stay engaged playing, for the game is not too easy that caused boredom or too hard that caused frustration to play \cite{AlexanderDDA}. 

\begin{figure}[h]
  \centering
  \includegraphics[width=0.7\columnwidth]{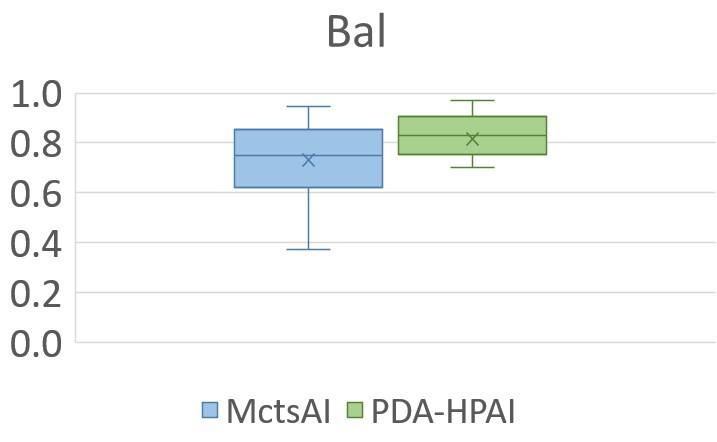}
  \caption{Comparison of the distribution of $Bal$ between fighting against the two AIs}
  \label{fig:BalancednessComparison}
\end{figure}

\begin{figure}[h]
  \centering
  \includegraphics[width=0.7\columnwidth]{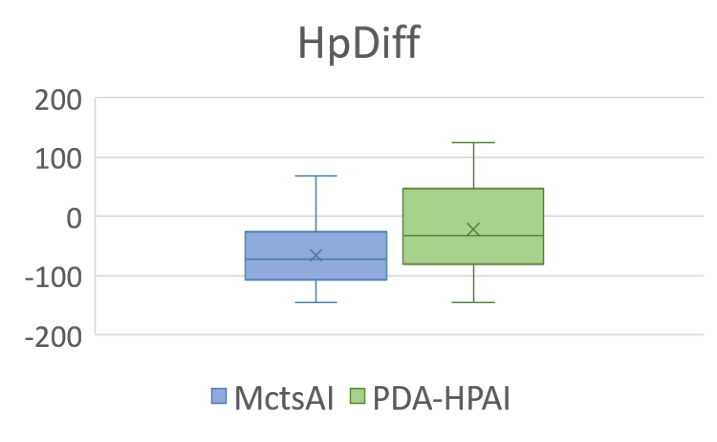}
  \caption{Comparison of the distribution of $HpDiff$ between fighting against the two AIs}
  \label{fig:HPDiffComparison}
\end{figure}

\clearpage

\begin{table}
\caption{Means with standard deviations of $Bal$ and $HpDiff$}\label{table1}
\centering
\begin{tabular}{|l|c|c|c|}
\hline
&  $Bal$ & $HpDiff$\\
\hline
MctsAI & $0.73\;\pm\;0.03$ & $-65.00\;\pm\;55.79$\\
PDA-HPAI & $0.82\;\pm\;0.23$ & $-21.89\;\pm\;82.83$\\
\hline
\end{tabular}
\end{table}

\subsubsection{Comparison with DDAHP-AI}
\begin{figure}[h]
  \centering
  \includegraphics[width=1\columnwidth]{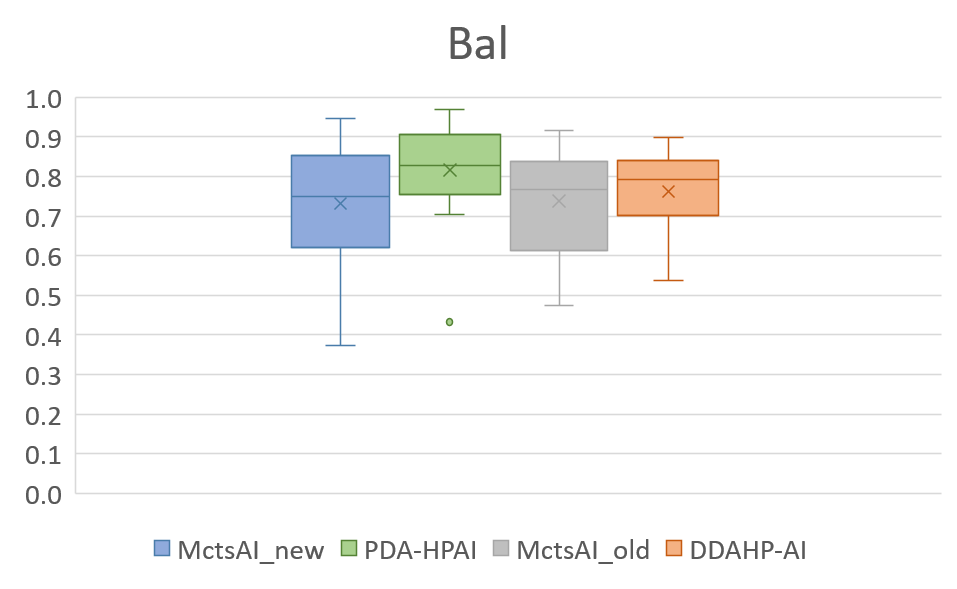}
  \caption{Comparison of the distribution of $Bal$ between fighting against (1) MctsAI in this study, (2) PDA-HPAI in this study, (3) MctsAI in the study by Kusano et al., and (4) DDAHP-AI in the study by Kusano et al.}
  \label{fig:ffff}
\end{figure}

\begin{figure}[h]
  \centering
  \includegraphics[width=0.8\columnwidth]{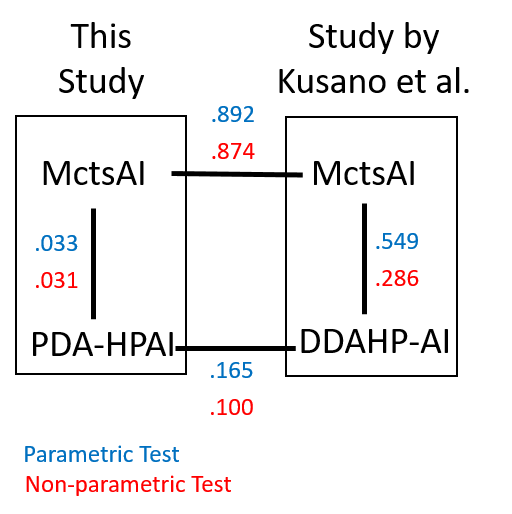}
  \caption{Statistical Differences in $Bal$}
  \label{fig:StatConclude}
\end{figure}

Finally, we compared $Bal$ between fighting against our proposed AI and DDAHP-AI by Kusano et al. \cite{kusanocog} (Fig. \ref{fig:ffff}), but no significant difference was found ($p$-value = .165 and .100 for independent sample t-test and Mann-Whitney U test respectively). Nevertheless, based on our findings that (1) $Bal$ against MctsAI in the two studies were not different ($p$-value = .892 and .874 for independent sample t-test and Mann-Whitney U test respectively), (2) $Bal$ against DDAHP-AI were not significantly different from against MctsAI in previous study ($p$-value = .549 and .286 for paired sample t-test and Wilcoxon signed-rank test respectively), and (3) PDA-HPAI is significantly better than MctsAI in this study, we conclude that PDA-HPAI outperformed DDAHP-AI. In addition, the mean $Bal$ against different AIs from Left to Right in Fig. \ref{fig:ffff}  were 0.73, 0.82, 0.74, and 0.76 respectively. These findings are summarized in Fig. \ref{fig:StatConclude}.

\subsection{Conclusions}
We have developed a system design and architecture for implementing a game AI that promotes health of the player. Result shows PDAHP-AI can recognize player's motion before player performs, and use it to analyze player's health state and determine actions that will encourage player do healthy motion that promote player's balance use of both sides of the body, which leads the more healthy way while performing actions during playing the motion game.

\chapter{Conclusions and Future Work}

Player Domination adjustment (PDA) is an novel proposed idea is that to control the AI's actions based on the player's inputs so as to adjust the player's dominant power. We proved its effectiveness that it leads to promotion of game-related self-efficacy. Stepping on this theory, several pieces of research on were conducted on a social deduction game and a fighting game respectively, show that using PDA is capable of promoting UX for the player. Also, PDA based AI outperforms previous AIs proposed by groups in our lab in two conducted experiments in terms of promoting health, enhancing $Bal$ in our conducted studies.

\section{Review of thesis contributions}

\begin{enumerate}
	\item Player Dominance Adjustment (PDA) \cite{myGCCE}, we proposed this novel idea, which is defined to control the AI's actions based on the player's inputs in the way that adjusts the player's dominant power. With the prediction of player's intention by considering player's behaviors/actions, we manipulate the game process in a way that follows the player's intentions and make them feel that they have the power to dominate the game or that game situations go in the way they expect. 
	
	\item the relationship among PDA, SE and UX \cite{myGCCE}, the first study using PDA, which implements on a social deduction game Werewolf was conducted. The result given by conducted experiment yields a prominent connection among self-efficacy, PDA and UX. This study provides an example for future research on using PDA for providing UX to the player.
	
	\item An opponent fighting-game AI based on PDA\cite{myMiG, myGPW} as well as its system design that stealthy promotes health of the player is proposed. In our system, this AI receives a player input motion in real time and data indicating use of body segments of the player from UKI, for determining its next action. The results demonstrated that the AI effectively encouraging them to use their body segments in a more balanced manner during motion gaming. 
	
\end{enumerate}

\section{Future work}
This thesis opens the door to new lines of game AI research. PDA can be used for several purposes, including providing UX and/or promoting health. Future work may investigate its effectiveness on better implementation as well as more effective methods for enhancing self-efficacy and experience of gameplay by adjusting dominant power. In addition, there is a promising direction for its performance on other senses of feedback, such as auditory or tactile in the research area of Human-Computer Interaction. The practicability of using PDA shown in proposed research shows its potential for making better player adaptive AI as well as the game system.


	\chapter*{List of publications}
	\addcontentsline{toc}{chapter}{List of publications}  
	
	{\setlength{\parindent}{0cm}
		\textbf{International Conference Proceedings}
	}
	\begin{itemize}
		
		\item \textbf{Xu, J.,} Paliyawan, P., Zhang, Y., Thawonmas, R. and Harada, T., 2019, October. Player Dominance Adjustment Motion Gaming AI for Health Promotion. In ACM SIGGRAPH Conference on Motion, Interaction and Games (p. 43). MIG ’19. Newcastle upon Tyne, United Kingdom: ACM.
		
		\item \textbf{Xu, J.,} Paliyawan, P., Thawonmas, R. and Harada, T., 2019, October. Player Dominance Adjustment: Promoting Self-Efficacy and Experience of Game Players by Adjusting Dominant Power. In 2019 IEEE 8th Global Conference on Consumer Electronics. GCCE 2019 (pp. 623–624) (Award Nominated Paper). Osaka, Japan: IEEE.
		
		\item \textbf{Xu, J., H,} Fang, Z., Ohno, S., Chen, Q., and Paliyawan, P., 2021, October. Fighting Game Commentator with Pitch and Loudness Adjustment Utilizing Highlight Cues. In The 10th global conference on consumer electronics. GCCE 2021 (pp. 673-677). Kyoto, Japan: IEEE.
		
		\item \textbf{Xu, J., H,} H, Cai,Y., Fang, Z., and Paliyawan, P, 2021, October. Promoting mental well-being for audiences in a live-streaming game by highlight-based bullet comments. In The 10th global conference on consumer electronics. GCCE 2021 (pp. 694–696). kyoto, Japan: IEEE.
		
	\end{itemize}
	
		{\setlength{\parindent}{0cm}
		\textbf{Domestic Conference Proceedings}
	}
	\begin{itemize}

	\item \textbf{Xu, J.,} Tatsuki, T., Fang, Z., Paliyawan, P., Harada, T., and Thawonmas, R., 2019, November. Player Adaptive Motion Gaming AI for Health Promotion (in Japanese). Game Programming Workshop 2019 (pp. 221–226). GPW-19. Hakone, Japan: IPSJ.

	\end{itemize}
	
\end{document}